\documentclass[aps,prl,twocolumn,superscriptaddress,groupedaddress]{revtex4}
\usepackage[compat=1.0.0]{tikz-feynman}
\usepackage{amssymb}
\usepackage{bm}
\usepackage{amsmath}
\usepackage{graphicx}
\usepackage{epstopdf}
\usepackage{natbib}
\usepackage{epsfig}
\usepackage{amsfonts}
\usepackage{mathrsfs}
\usepackage[toc,page,title,titletoc,header]{appendix}
\usepackage[colorlinks,linkcolor=blue,citecolor=blue,anchorcolor=blue]{hyperref}
\usepackage{dsfont,amsthm,amsbsy}
\usepackage{physics}
\usepackage{floatrow}

\usepackage{fancyhdr}
\usepackage{ulem}
\usepackage{bbold}
\usepackage{tikz}

\usepackage{bold-extra}
\newcommand{\inlinesection}[1]{
    \addcontentsline{toc}{section}{#1}
    \noindent{\bfseries #1}}
\usepackage{array}
\newcolumntype{L}[1]{>{\raggedright\let\newline\\\arraybackslash\hspace{0pt}}m{#1}}
\newcolumntype{C}[1]{>{\centering\let\newline\\\arraybackslash\hspace{0pt}}m{#1}}
\newcolumntype{R}[1]{>{\raggedleft\let\newline\\\arraybackslash\hspace{0pt}}m{#1}}

\begin{document}
\title{Enhanced pair-density-wave vertices in a bilayer Hubbard model at half-filling}
\author{Fangze Liu}
\affiliation{Department of Physics, Stanford University, Stanford, California 94305, USA}
\affiliation{Stanford Institute for Materials and Energy Sciences, SLAC National Accelerator Laboratory, 2575 Sand Hill Road, Menlo Park, California 94025, USA}

\author{Xu-Xin Huang}
\affiliation{Stanford Institute for Materials and Energy Sciences, SLAC National Accelerator Laboratory, 2575 Sand Hill Road, Menlo Park, California 94025, USA}
\affiliation{Department of Applied Physics, Stanford University, Stanford, California 94305, USA}

\author{Edwin W. Huang}
\affiliation{Department of Physics and Institute of Condensed Matter Theory, University of Illinois at Urbana-Champaign, Urbana, Illinois 61801, USA}
\affiliation{Department of Physics and Astronomy, University of Notre Dame, Notre Dame, Indiana 46556, United States}
\affiliation{Stavropoulos Center for Complex Quantum Matter, University of Notre Dame, Notre Dame, Indiana 46556, United States}

\author{Brian Moritz}
\affiliation{Stanford Institute for Materials and Energy Sciences, SLAC National Accelerator Laboratory, 2575 Sand Hill Road, Menlo Park, California 94025, USA}

\author{Thomas P. Devereaux}
\affiliation{Stanford Institute for Materials and Energy Sciences, SLAC National Accelerator Laboratory, 2575 Sand Hill Road, Menlo Park, California 94025, USA}
\affiliation{Department of Materials Science and Engineering, Stanford University, Stanford, California 94305, USA}

\begin{abstract}
Motivated by the pair-density-wave (PDW) state found in the one-dimensional Kondo-Heisenberg chain, we report on a determinant quantum Monte Carlo DQMC study of pair-fields for a two-dimensional half-filled Hubbard layer coupled to an itinerant, non-interacting layer with one electron per site.  In a specific range of interlayer hopping, the pairing vertex associated with PDW order becomes more attractive than that for uniform $d$-wave pairing, although both remain subdominant to the leading antiferromagnetic correlations at half-filling. Our result sheds light on where one potentially may find a PDW state in such a model.
\end{abstract}

\maketitle
A novel spatially modulated superconducting state known as a pair-density wave (PDW)~\cite{agterberg2020physics} has attracted increasingly significant attention. This nonuniform unidirectional superconducting state was initially proposed to provide a phenomenological understanding of the high-temperature cuprate superconductors~\cite{PhysRevLett.99.127003, PhysRevB.79.064515}, and experimental evidence supporting its existence has been observed in a wide range of quantum materials, including cuprates~\cite{rajasekaran2018probing, ruan2018visualization, edkins2019magnetic, choubey2020atomic, du2020imaging, shi2020pair, PhysRevX.11.011007, wang2021scattering}, heavy fermion materials~\cite{gu2023detection, aishwarya2023magnetic}, iron-based superconductors~\cite{zhao2023smectic, liu2023pair}, kagome superconductors~\cite{chen2021roton}, and transition-metal dichalcogenides~\cite{liu2021discovery}. 
Numerically, it has been first observed in a density-matrix renormalization group (DMRG) study~\cite{PhysRevLett.105.146403} of a one-dimensional Kondo-Heisenberg chain, which consists of a Heisenberg antiferromagnetic (AFM) chain coupled to a free electron gas. More recently, the PDW order has been found in several other quasi-1D models with a quasi-long-range and divergent PDW susceptibility through DMRG~\cite{huang2022pair, PhysRevB.107.214504, hcjiang2023pair, yfjiang2023pair}, in 2D microscopic models with PDW as the leading divergent susceptibility by using renormalization group analysis~\cite{PhysRevLett.130.126001, PhysRevLett.131.026601, PhysRevB.108.035135}, and in certain exactly solvable models~\cite{han2024quantum, PhysRevLett.129.177601, PhysRevB.108.174506}.

In this Letter, we employ the numerically exact determinant quantum Monte Carlo (DQMC) approach~\cite{PhysRevB.40.506, PhysRevD.24.2278} to search for fingerprints of the PDW in a two-dimensional system. Our model comprises a bilayer structure in two dimensions, consisting of a non-interacting layer and a repulsive Hubbard layer with strong electronic correlations. The correlation is characterized by the ratio of the on-site interaction to intralayer nearest-neighbor hopping, denoted as $U/t$. The layers are hybridized through an interlayer tunneling, also referred to as the hopping matrix element $t_{\perp}$. 
In the strong-coupling limit, the half-filled Hubbard layer maps to an insulating Heisenberg antiferromagnet; and in this context, our model effectively reduces to the two-dimensional Kondo-Heisenberg model, characterized by exchange interactions $J_{\text{K}}\sim 2 t_{\perp}^2/U$ and $J_{\text{H}} \sim 4 t^2/U$, which may hold promise as a candidate for PDWs in two dimensions.

The primary focus of our study is to investigate pairing tendencies of this bilayer model in the special case of one electron per site in both layers. As our model is defined on a bipartite lattice, the imposed condition of one electron per site (half-filling) in both layers leads to a sign-problem-free implementation of DQMC~\cite{li2019sign}, allowing us to access relatively low temperatures. 
At half-filling, pair-field correlations remain subdominant to the AFM spin-spin correlations driven by the strong Coulomb repulsion within the Hubbard layer. However, our findings reveal strong subdominant pair-field susceptibility with leading PDW singlet pairing vertices at the center-of-mass momentum $\mathbf{q}=(\pi,\pi)$, particularly when the interlayer hybridization reaches a regime where $J_{\text{K}} \sim J_{\text{H}}$. 
Previous DQMC studies~\cite{PhysRevLett.108.246401, PhysRevB.88.235123, PhysRevB.91.155119} have explored the magnetic and transport properties of this model at half-filling, highlighting the competition between AFM and Kondo-singlet formation at intermediate hybridization --- supported also here, as detailed in the Supplemental Material (SM). This suggests that the competition between intralayer AFM and interlayer local spin-singlet formation provides conditions favorable for PDW formation. 
Despite the significant challenges posed by the fermion sign problem in DQMC simulations, we also explore the PDW pair-fields when doping the model away from half-filling, as discussed in the final section.

\inlinesection{Model and Method.}
In this Letter, we consider a two-dimensional bilayer Hubbard Hamiltonian,
\begin{equation}
\begin{aligned}
\hat{H} 
& = - \mathop{\sum\limits_{\langle ij \rangle \sigma}}_{\ell \in \{A,B\}} t_{\ell} (\hat{c}_{i \ell \sigma}^{\dagger} \hat{c}_{j \ell \sigma} + \hat{c}_{j \ell \sigma}^{\dagger} \hat{c}_{i \ell \sigma}) \\
& \quad - \sum\limits_{i \sigma} t_{\perp} (\hat{c}_{i A \sigma}^{\dagger} \hat{c}_{i B \sigma} +\hat{c}_{i B \sigma}^{\dagger} \hat{c}_{i A \sigma}) \\
& \quad + \mathop{\sum\limits_{i} U_{\ell}}_{\ell \in \{A,B\}} (\hat{n}_{i \ell \uparrow} - \frac{1}{2})(\hat{n}_{i \ell \downarrow} - \frac{1}{2}) \\
& \quad - \mathop{\sum\limits_{i \sigma} \mu_{\ell}}_{\ell \in \{A,B\}} \hat{n}_{i \ell \sigma},
\end{aligned}
\end{equation}
where $\hat{c}_{i \ell \sigma}^{\dagger}$ ($\hat{c}_{i \ell \sigma}$) denotes the creation (annihilation) operator of an electron with spin $\sigma$ at site $i$ and layer $\ell$, and $\hat{n}_{i \ell \sigma} = \hat{c}_{i \ell \sigma}^{\dagger} \hat{c}_{i \ell \sigma}$ is the corresponding electron density. 
$U_{\ell}$ is the local (on-site) Coulomb interaction. We set $U_A/t = 8$, indicating a repulsive on-site Hubbard interaction in layer $A$, and $U_B/t = 0$, making layer $B$ non-interacting. $\mu_{\ell}$ is the chemical potential, which tunes the electron density in layer $\ell$. 
$t_{\ell}$ denotes the hopping integral between nearest-neighbor sites in the same layer $\ell$, and $t_{\perp}$ is the hopping integral between layers, which controls the interlayer hybridization. We set $t_{\ell}=t=1$ in both layers and measure all energies in units of $t$. In this Letter, we focus on a range of $t_{\perp}$ such that $J_{\text{K}} \lesssim J_{\text{H}}$.

We use the numerically exact DQMC algorithm to simulate the model at finite temperatures on square lattices with periodic boundary conditions (see the SM for a brief discussion of the Trotter-Suzuki decomposition and discrete Hubbard-Stratonovich transformation, as well as Ref.~\cite{PhysRevB.40.506} for methodological detail). 
Our primary focus is on half-filling in both layers, utilizing a particle-hole symmetric band structure to avoid the fermion sign problem~\cite{PhysRevB.92.045110}. For this purpose, we perform simulations on $10 \times 10$ clusters down to relatively low temperatures ($T = 1/\beta = t/30$). To avoid Trotter error at low temperatures, we choose $\Delta\tau \approx 0.01 = 0.8/\sqrt{U W}$, which is well below the conventional limit of $1/\sqrt{U W}$, where $U$ is the interaction strength and $W$ is the non-interacting bandwidth. 
In the final section, we investigate the effect of hole doping, but at higher temperatures ($T=1/8$) and on smaller $8\times 8$ clusters to mitigate challenges posed by the sign problem. \\


\begin{figure}[h!]
\includegraphics[width=1\linewidth]{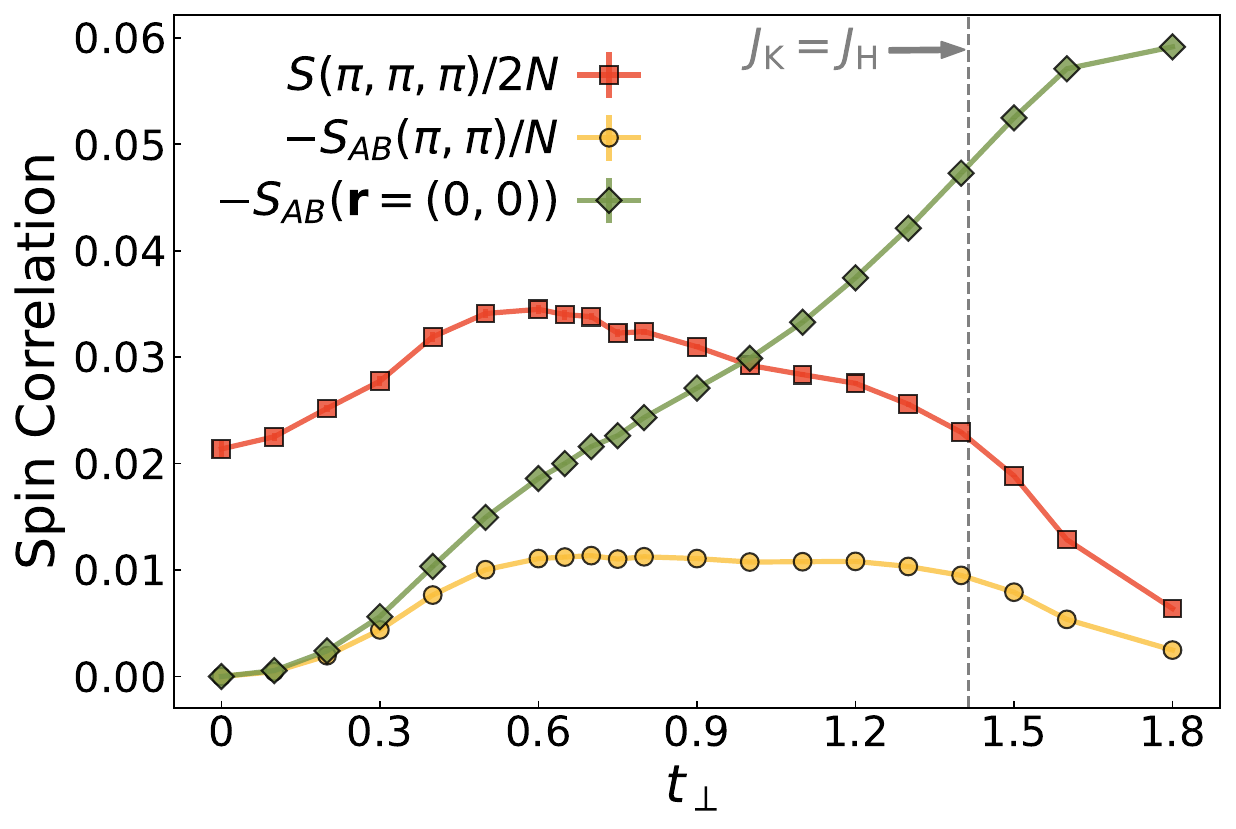}
\caption{\label{ss}
\textbf{Equal-time spin-spin correlations.}
The $t_{\perp}$ dependence of the spin-spin correlation at the AFM wavevector $S(\pi,\pi,\pi)$, the interlayer AFM spin correlation $S_{AB}(\pi,\pi)$ and interlayer local spin correlation $S_{AB}(\mathbf{r}=(0,0))$ for the half-filled system ($n_A=n_B=1$) of size $N = 10 \times 10$ with $\beta=30$.
}
\end{figure}

\inlinesection{Spin Correlations.}
Before investigating the pair-field correlations within the model, we first analyze the spin-spin correlations, which are unsurprisingly the dominant correlations at half-filling. 

The real space, equal-time spin-spin correlation function is defined as 
\begin{equation}
\begin{aligned}
S_{\ell \ell^{\prime}}(\mathbf{r}) 
&= \langle \hat{s}^z_{r\ell} \hat{s}^z_{0\ell^{\prime}} \rangle
\end{aligned}
\end{equation}
with $\hat{s}^z_{i \ell}= (\hat{n}_{i \ell \uparrow}-\hat{n}_{i \ell \downarrow})/2$, and its Fourier transform gives the structure factor
$S_{\ell \ell^{\prime}}(\mathbf{q}) = \sum_{\mathbf{r}} e^{i \mathbf{q} \cdot \mathbf{r}} S_{\ell \ell^{\prime}}(\mathbf{r})$.
We will consider interlayer bonding $\left(\hat{s}^z_{r A}+\hat{s}^z_{r B}\right)$ and anti-bonding $\left(\hat{s}^z_{r A}-\hat{s}^z_{r B}\right)$ combinations, which are used to define a ``$q_z$" quasi-momentum of $0$ or $\pi$, respectively.

In the absence of interlayer hybridization, the equal-time spin-spin correlation function peaks at the in-plane AFM wave vector $\mathbf{q}=(\pi,\pi)$, indicative of strong AFM correlations within the Hubbard layer. 
After turning on the interlayer hybridization, the spin-spin correlation function is dominated by the anti-bonding $\mathbf{q}=(\pi,\pi,\pi)$ wavevector. Moreover, the interlayer spin correlations $S_{AB}(\mathbf{q})$ exhibit a peak at the in-plane AFM wavevector $\mathbf{q}=(\pi,\pi)$. Figure~\ref{ss} shows that for small values of $t_{\perp}$, both the spin-spin correlations at $\mathbf{q}=(\pi,\pi,\pi)$ and the long-range interlayer AFM spin correlations initially increase with $t_{\perp}$, a behavior attributed to the development of superexchange interaction across the layers. 
Notably, the interlayer hybridization opens a gap for the itinerant electrons, resulting in a transition to an insulating phase. This phenomenon was previously reported in Ref.~\cite{PhysRevB.88.235123} and is further substantiated in the SM. 
As $t_{\perp}$ further increases, both $S(\pi,\pi,\pi)$ and $-S_{AB}(\pi,\pi)$ start to drop quickly around $t_{\perp} \sim \sqrt{2}$. In this regime, neither the effective interlayer Kondo coupling $J_{\text{K}} \sim 2 t_{\perp}^2/U$, nor the intralayer Heisenberg coupling $J_{\text{H}} \sim 4 t^2/U$ dominates; however, the interlayer local spin-spin correlation function $S_{AB}\big(\mathbf{r}=(0,0)\big)$ increases rapidly and appears to saturate as as $t_{\perp}$ grows. The behavior of the spin-spin correlations suggests a strong magnetic competition between intralayer AFM and interlayer singlet formation across $t_{\perp} \sim \sqrt{2}$. \\


\begin{figure}[htb]
\includegraphics[width=1\linewidth]{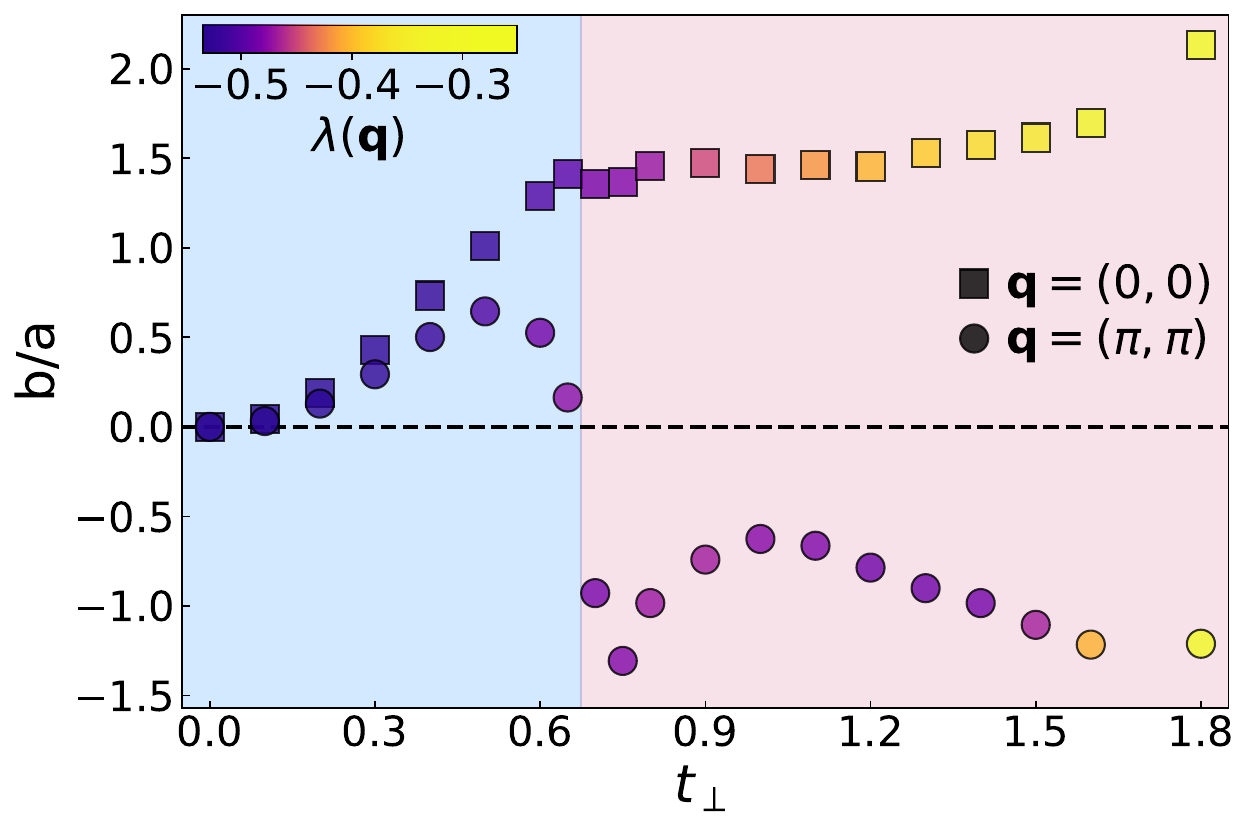}
\caption{\label{pair_eigenvec}
\textbf{Eigenvectors of $\Gamma(\mathbf{q}) \bar{P}(\mathbf{q})$.}
The left eigenvector corresponding to the largest negative eigenvalues of $\Gamma(\mathbf{q}) \bar{P}(\mathbf{q})$, denoted as $\mathbf{\phi}(\mathbf{q})=$ $(a,-a,b,-b,0)$ in the operator basis $\left(\hat{\varsigma}_{\mathbf{q},x,A}, \hat{\varsigma}_{\mathbf{q},y,A}, \hat{\varsigma}_{\mathbf{q},x,B}, \hat{\varsigma}_{\mathbf{q},y,B}, \hat{\varsigma}_{\mathbf{q},z} \right)$, is represented by the ratio $a/b$ on the y axis. The figure includes the center-of-mass momentum $\mathbf{q}=(0,0)$ (square symbols) and $\mathbf{q}=(\pi,\pi)$ (circle symbols), with the symbol color indicating the magnitude of the largest negative eigenvalue, $\lambda(\mathbf{q})$. Blue region denote where $\lambda(0,0)$ is the largest negative eigenvalue across all considered $\mathbf{q}$ values, while red signifies $\lambda(\pi,\pi)$ as the largest negative. Simulations are conducted on a half-filled $10 \times 10$ system ($n_A=n_B=1$) at $\beta=30$.
}
\end{figure}

\begin{figure}[htb] 
\includegraphics[width=1\linewidth]{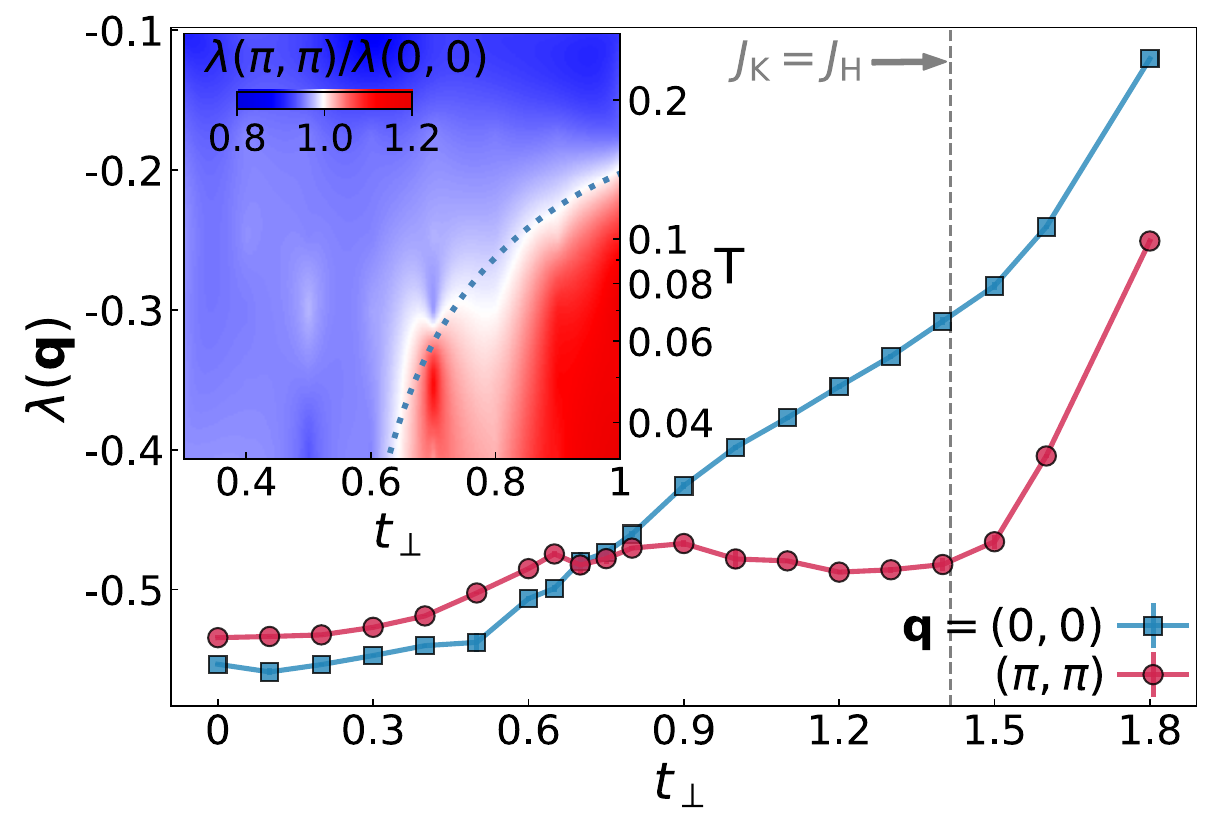}
\caption{\label{pair_eigen}
\textbf{Largest negative eigenvalues of $\Gamma(\mathbf{q}) \bar{P}(\mathbf{q})$.} 
$\lambda(\mathbf{q})$ for $\mathbf{q}=(0,0)$ and $(\pi,\pi)$ as a function of $t_{\perp}$ at $\beta=30$ on a $10\times 10$ cluster. Errors estimates, smaller than the data points, are obtained by bootstrap resampling.
Inset: A color map of $\lambda(\pi,\pi)/\lambda(0,0)$ across different temperature $T$ and $t_{\perp}$. The blue region indicates $\lambda(0,0) >\lambda(\pi,\pi)$, while the red region where $\lambda(0,0) < \lambda(\pi,\pi)$ highlights the dominance of a PDW superconducting instability at $\mathbf{q}=(\pi,\pi)$. 
The blue dashed line is a guide-to-the-eye in the crossover region.}
\end{figure}

\begin{figure}[h!] 
\includegraphics[width=1\linewidth]{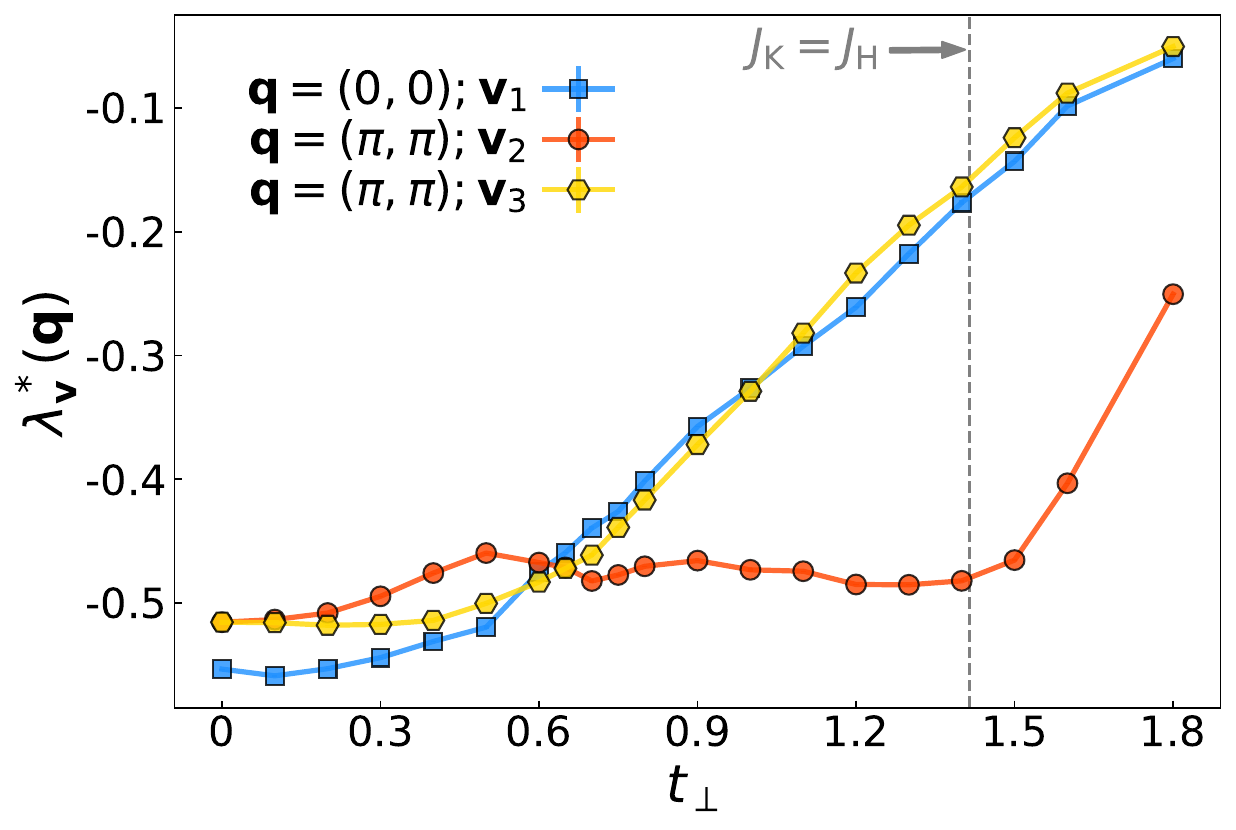}
\caption{\label{pair_vec}
\textbf{Pairing estimator for trial pair-field vectors.}
$\lambda^*_{\mathbf{v}}$ for various trial wavevectors $\mathbf{v}$, at different center-of-mass momentum $\mathbf{q}$. Considered vectors: $\mathbf{v}_1=(1,-1,0,0,0)/\sqrt{2}$ (blue), $\mathbf{v}_2=(1,-1,-1,1,0)/2$ (orange), $\mathbf{v}_3=(1,-1,1,-1,0)/2$ (yellow). Measurements were performed on a $N=10\times10$ cluster at $\beta=30$.}
\end{figure}

\begin{figure*}[thb] 
\centering
\begin{minipage}{0.8\textwidth}
    \text{(a) Hole doping in layer $B$}
    \includegraphics[width=\linewidth]{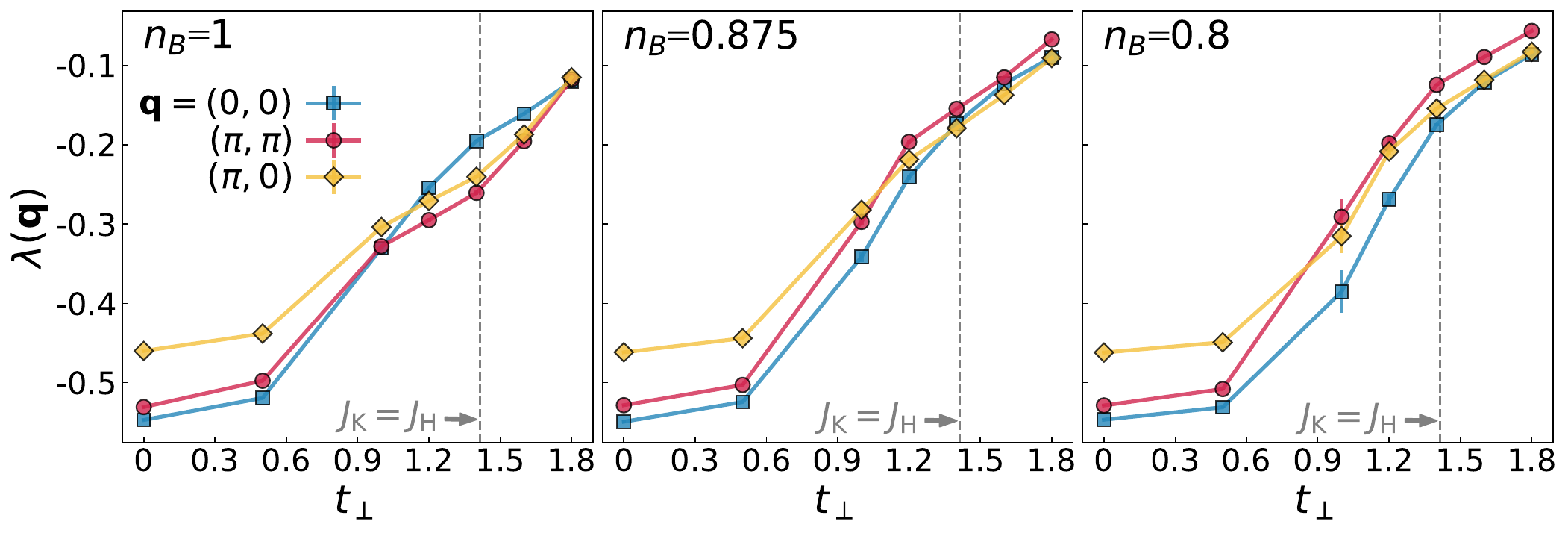} 
\end{minipage}\\
\begin{minipage}{0.8\textwidth}
    \text{(b) Hole doping in layer $A$}
    \includegraphics[width=\linewidth]{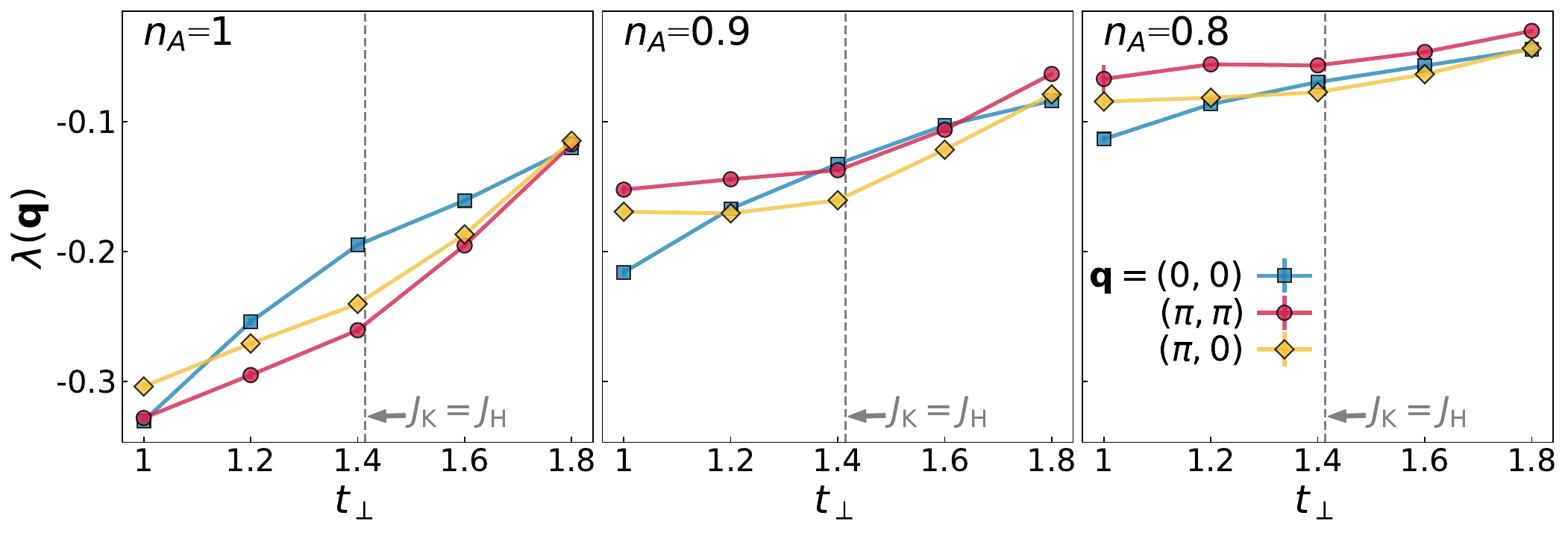} 
\end{minipage}\\
\caption{
\textbf{Comparison between half-filling and hole dopings} The largest negative eigenvalues of $\Gamma(\mathbf{q}) \bar{P}(\mathbf{q})$ as a function of $t_{\perp}$ at different $\mathbf{q}$ 
when (a) $n_B=1, 0.875, 0.8$, $n_A=1 \pm 0.004$ and (b) $n_A=1, 0.9, 0.8$, $n_B=1 \pm 0.03$. Simulations are done at $\beta=8$ on a $8\times 8$ cluster. Errors, smaller than the data points, were estimated by bootstrap resampling.}
\label{pair_eigen_doping}
\end{figure*}

\inlinesection{Pair-Fields at Half-Filling.}
While the half-filled Hubbard model has dominant AFM spin correlations, various analytical and numerical studies suggest that the leading superconducting instability near half-filling corresponds to a uniform spin-singlet $d_{x^2-y^2}$-wave pair-field~\cite{Arovas_2022}. Following the discovery of a quasi-long-range PDW state on the Kondo-Heisenberg chain, we confine our attention to exploring how coupling with itinerant electrons influences the superconducting pair-fields, particularly at nonzero center-of-mass momentum.

To describe singlet pair-fields, we define the operator 
\begin{equation}
\hat{\mathbf{\Delta}}(\mathbf{q}) \\
= \left(\hat{\varsigma}_{\mathbf{q},x,A}, \hat{\varsigma}_{\mathbf{q},y,A}, \hat{\varsigma}_{\mathbf{q},x,B}, \hat{\varsigma}_{\mathbf{q},y,B}, \hat{\varsigma}_{\mathbf{q},z} \right), 
\end{equation}
as a function of the center-of-mass momentum $\mathbf{q}$.
Here, $\hat{\varsigma}_{\mathbf{q},\alpha,\ell} = \sum_i e^{i \mathbf{q}\cdot \mathbf{r}_i} \left(\hat{c}_{(i+\alpha) \ell \uparrow}\hat{c}_{i \ell \downarrow}-\hat{c}_{(i+\alpha) \ell \downarrow} \hat{c}_{i \ell \uparrow}\right)/2$ denotes the singlet pair operator on the nearest-neighbor bond $\alpha$ in layer $\ell$, and $\hat{\varsigma}_{\mathbf{q},z} =\sum_i e^{i \mathbf{q}\cdot \mathbf{r}_i} \left(\hat{c}_{i A \uparrow} \hat{c}_{i B \downarrow}-\hat{c}_{i A \downarrow} \hat{c}_{i B \uparrow}\right)/2$ indicates interlayer pairing. 
The full pair-field susceptibility is a square matrix written as
\begin{equation}
P (\mathbf{q}) = \int_0^{\beta} d\tau \langle \hat{\mathbf{\Delta}}(\tau, \mathbf{q}) \hat{\mathbf{\Delta}}^{\dagger} (\mathbf{q}) \rangle.
\end{equation}
Similarly, we can calculate the uncorrelated pair-field susceptibility $\bar{P}(\mathbf{q})$~\cite{PhysRevB.39.839}, which is the disconnected part of $P (\mathbf{q})$ (see the expressions given in the SM).

To elucidate the dominant pairing interaction within the system, we compute the pairing vertex $\Gamma$ using the Dyson-like expression
$\Gamma(\mathbf{q}) = P^{-1}(\mathbf{q}) - \bar{P}^{-1}(\mathbf{q})$.
Equivalently, the full pair-field susceptibility takes the form
\begin{equation}\label{eq:P}
P(\mathbf{q}) = \bar{P}(\mathbf{q}) \left(\mathbb{1} + \Gamma(\mathbf{q}) \bar{P}(\mathbf{q})\right)^{-1}.
\end{equation}
From this expression, $\Gamma(\mathbf{q}) \bar{P}(\mathbf{q})$ can be used to gauge pairing tendencies: the pairing mode with the most divergent susceptibility is identified through the eigenmode of $\Gamma(\mathbf{q}) \bar{P}(\mathbf{q})$ corresponding to the eigenvalue closest to $-1$.

In situations where numerical analysis of the $\mathbf{q}$-dependent equal-time pair-field correlations in our model at half-filling does not provide direct evidence for a dominant PDW state down to the lowest accessible temperatures, we explore the leading pairing tendencies through $\Gamma(\mathbf{q}) \bar{P}(\mathbf{q})$. 
Figure~\ref{pair_eigenvec} shows the evolution of the largest negative eigenvalues $\lambda(\mathbf{q})$ of $\Gamma(\mathbf{q}) \bar{P}(\mathbf{q})$ with interlayer hopping $t_{\perp}$, alongside the corresponding center-of-mass momentum $\mathbf{q}$, and its left eigenvector $\mathbf{\phi}(\mathbf{q})$. When $\lambda(\mathbf{q})\rightarrow -1$, this eigenvector corresponds to the most divergent eigenmode of $P(\mathbf{q})$ (as discussed in the SM). 

We find that the left eigenvector corresponding to the largest negative eigenvalues of this matrix takes the form $\mathbf{\phi}(\mathbf{q})=(a,-a,b,-b,0)$ at all $t_{\perp}$ ($a,b\in \mathbb{R}$). 
At small $t_{\perp}$, the leading pairing instability corresponds to a uniform $d$-wave order parameter with pairing field mainly in layer $A$, and the pair-fields on the two layers have the same phase. As $t_{\perp}$ increases to $0.7$ or above, the leading instability changes to a PDW with an in-plane center-of-mass momentum $\mathbf{q}=(\pi,\pi)$. The corresponding pairing order parameter takes a $d$-wave form $|a|(\hat{\varsigma}_{\mathbf{q},x,A}-\hat{\varsigma}_{\mathbf{q},y,A})-|b|(\hat{\varsigma}_{\mathbf{q},x,B}-\hat{\varsigma}_{\mathbf{q},y,B})$, where $a$ and $b$ have comparable amplitudes but opposite signs. 
Within numerical accuracy, these leading eigenstates are degenerate with another set of eigenstates, which are characterized by eigenvectors that adopt an extended $s$-wave form $|a|(\hat{\varsigma}_{\mathbf{q},x,A}+\hat{\varsigma}_{\mathbf{q},y,A})-|b|(\hat{\varsigma}_{\mathbf{q},x,B}+\hat{\varsigma}_{\mathbf{q},y,B})$, also with a center-of-mass momentum of $\mathbf{q}=(\pi,\pi)$. This degeneracy is attributed to symmetry at $\mathbf{q}=(\pi,\pi)$~\footnote{After Fourier transforming the order parameter at $\mathbf{q}=(\pi,\pi)$ with the $d$-wave form and that with the extended $s$-wave form into real space, their patterns are related by the reflection symmetry along y-axis}.

We trace the evolution of pairing tendencies at both $\mathbf{q}=(0,0)$ and $(\pi,\pi)$ and show the largest negative eigenvalue of $\Gamma(\mathbf{q}) \bar{P}(\mathbf{q})$, $\lambda(\mathbf{q})$, as a function of $t_{\perp}$ for both momentum points in Fig.~\ref{pair_eigen}. A crossover from uniform pairing to a PDW is observed around $t_{\perp}=0.7$. The inset illustrates the ratio $\lambda(\pi,\pi)/\lambda(0,0)$ as a function of temperature $T$ and $t_{\perp}$ to highlight the crossover region. 
For $t_{\perp} > 0.7$, $\lambda(\pi,\pi)$ exhibits a broad minimum, roughly centered around an intermediate range of $t_{\perp}$ where $J_{\text{K}}\sim J_{\text{H}}$. Within this range, the gap between $\lambda(\pi,\pi)$ and $\lambda(0,0)$ expands as $t_{\perp}$ increases (see the SM for additional details).

The leading eigenvectors change considerably as a function of model parameters in the crossover region. To gain insight into the pairing structure, we compute a pairing estimator, $\lambda^*_{\mathbf{v}}(\mathbf{q}) = \left(\mathbf{v}^{T} \Gamma(\mathbf{q}) \mathbf{v})/(\mathbf{v}^{T} \bar{P}(\mathbf{q})^{-1} \mathbf{v}\right)$, for representative trial vectors $\mathbf{v}$ in the basis $\left(\hat{\varsigma}_{\mathbf{q},x,A}, \hat{\varsigma}_{\mathbf{q},y,A}, \hat{\varsigma}_{\mathbf{q},x,B}, \hat{\varsigma}_{\mathbf{q},y,B}, \hat{\varsigma}_{\mathbf{q},z}\right)$. Detailed methodology is described in the SM. 
As demonstrated in both Fig.~\ref{pair_eigenvec} and Fig.~\ref{pair_vec}, the leading pairing eigenmode for $t_{\perp}<0.7$ has an eigenvector that represents uniform $d$-wave pair fields in layer $A$. Up to the crossover value of $t_{\perp}$, the values of $\lambda^*_{\mathbf{v}}(\mathbf{q})$ for the two PDW trial vectors are comparable --- one with in-phase (yellow curve) and the other with antiphase (orange curve) pair-field structure on two layers. However, through the crossover region, a considerable splitting occurs between the in-phase and antiphase PDWs, and the antiphase PDW at $\mathbf{q}=(\pi,\pi)$ markedly becomes the most dominant pairing tendency. \\

\inlinesection{Pair-Fields upon Doping.} 
Although our results showing an indication of strong PDW correlations are encouraging, they also reveal that the strength of the superconducting correlations at overall half filling is insufficient to overcome antiferromagnetic correlations. It is anticipated that doping the system away from half filling could create a spin gap, suppress AFM correlations, and thereby enhance PDW correlations. To investigate this hypothesis, we will further explore the performance of our model in the underdoped regime in this section. 

Since the PDW in the 1-D Kondo-Heisenberg model was discovered at $1/8$ hole doping in the electron gas, we naturally start with doping layer $B$. 
Figure~\ref{pair_eigen_doping}a presents a comparison of the largest negative eigenvalues of $\Gamma(\mathbf{q}) \bar{P}(\mathbf{q})$ for both half-filling and the hole doped scenarios ($n_B=0.875, 0.8$), with layer $A$ maintained at half-filling, but at an elevated temperature ($\beta=8$) compared to the previous sections, which is constrained by the fermion sign problem~\footnote{For the doped cases shown in Fig.~\ref{pair_eigen_doping}, when $n_B = 0.8$, the sign ranges from $0.0632$ ($t_{\perp} = 1$) to $0.905$ ($t_{\perp} = 1.8$). For the other doping levels, the sign generally ranges from $\sim 0.3$ ($t_{\perp} = 1$) to $\sim 0.9$ ($t_{\perp} = 1.8$).  However, at a lower temperature ($\beta = 12$), the sign rapidly drops to $0.001$, and for a larger system size ($10\times 10$), it decreases further to $0.0001$}. 
In the regime where $J_{\text{K}} \sim J_{\text{H}}$, the uniform $d$-wave and PDW pair-fields at $\mathbf{q}=(\pi,\pi)$ and $(\pi,0)$ have nearly identical eigenvalues in the doped cases, despite the fact that AFM correlations have been suppressed by the introduction of carriers (see the SM). It is possible that the limitations imposed by the fermion sign problem prevent us from observing a notable increase in the magnitude of PDW eigenvalues upon hole doping, but it also could be that the robustness of AFM correlations leaves no room for the PDW to prevail. What happens if we directly disrupt the AFM correlations by doping holes into layer $A$? 

We subsequently investigate the impact of the hole doping in layer $A$ on PDWs, as depicted in Fig.~\ref{pair_eigen_doping}b. This analysis focuses on relatively large values of the interlayer hopping to reduce the impact of the sign problem. Upon doping layer $A$, the AFM correlations are destroyed and the local interlayer spin correlations dominate the system, as detailed in the SM. Meanwhile, we observe that $\lambda(\mathbf{q})$ becomes less negative across all center-of-mass momenta $\mathbf{q}$, and notably the leading pairing eigenvalues are located at $\mathbf{q}=(\pi,0)$ and $(0,\pi)$, particularly when $J_{\text{K}}\sim J_{\text{H}}$. We find that the spin response peaks at $\mathbf{q}=(\pi,\pi)$, with secondary peaks at $\mathbf{q}=(\pi,0)$ and $(0,\pi)$ when the system is lightly doped (see the SM). These findings suggest a potential association between the PDW wavevector and local magnetic ordering wavevectors. 

For the parameters and temperatures we can access in the doped cases, our results cannot be over-interpreted as to the ultimate leading eigenmode in the pairing channel. Nevertheless, our results are suggestive of the dominant pairing wavevector $(\pi,0)$ (or $(0,\pi)$) near $t_{\perp} \sim \sqrt{2}$ upon doping. 
The observed modulation of a PDW could potentially be rationalized by the inter-Fermi-pocket nesting from the perspective of weak-coupling, which might be similar to the phenomena observed in other systems~\cite{PhysRevLett.130.126001, PhysRevLett.131.026601, PhysRevB.107.224516}. 
However, due to the momentum resolution and lack of sharp Fermi surfaces, rigorously analyzing the connection between our observations and Fermi surface nesting is challenging. Nonetheless, we provide measurements of the Fermi surface in SM for future reference. From our current analysis, the PDW tendency upon doping might be explained more explicitly from a strong-coupling perspective~\cite{PhysRevB.109.L121101}, which suggests that ``wrong sign'' pair-hopping terms in the effective model might be responsible for the exotic pairing possibilities.

Despite attempts to induce PDW states by doping holes into the systems, evidence for the existence of PDWs in our model remains elusive. Beyond exploring lower temperatures and fine-tuning model parameters, alternative approaches such as modifying the hopping terms, as suggested in Ref.~\cite{PhysRevB.107.214504}, or incorporating additional interaction terms as in Ref. \cite{yfjiang2023pair}, might be potential pathways for achieving such states.\\

\inlinesection{Conclusion.}
In this Letter, we investigate the pairing tendencies of a half-filled 2D repulsive Hubbard model hybridized to non-interacting electrons. This model system exhibits pronounced AFM tendencies when the interlayer hybridization, $t_{\perp}$, satisfies $J_{\text{K}} \lesssim J_{\text{H}}$.
For values of $t_{\perp}$ where $J_{\text{K}} \gtrsim J_{\text{H}}$, in-plane AFM correlations are suppressed, accompanied by an increase in local interlayer spin singlet correlations. 
In between these two extremes, as $t_{\perp}$ increases, we identify a crossover in pair-field modes from a uniform $d$-wave form to a PDW at a center-of-mass momentum $(\pi,\pi)$, and the PDW tendency becomes most significant when $J_{\text{K}} \sim J_{\text{H}}$. This suggests that a regime where the effective Kondo coupling is comparable to the effective Heisenberg coupling fosters a strong competition between intralayer AFM and interlayer singlet formation, which potentially constitute ideal conditions for the emergence of a PDW superconducting state. Our study can serve as an invitation for further exploration into this topic in unexplored parameter regimes and associated models.\\

\inlinesection{Acknowledgment.}
F.L. acknowledges the helpful discussion with Zhaoyu Han on pair-density waves. This work was supported by the U.S. Department of Energy (DOE), Office of Basic Energy Sciences, Division of Materials Sciences and Engineering under Contract No. DE-AC02-76SF00515. E.W.H. was supported by the Gordon and Betty Moore Foundation’s EPiQS Initiative through grants GBMF 4305 and GBMF 8691 at the University of Illinois Urbana-Champaign. Computational work was performed on the Sherlock cluster at Stanford University and on resources of the National Energy Research Scientific Computing Center, supported by the U.S. DOE, Office of Science, under Contract no. DE-AC02-05CH11231.

\inlinesection{Code and data availability}
Our DQMC simulation code will be available at \footnote{https://github.com/edwnh/dqmc}. Data supporting this Letter are stored on the Sherlock cluster at Stanford University and are available upon request.\\

\bibliography{ref}

\end{document}


\title{\Large Supplemental Materials\\
\large Enhanced pair-density-wave vertices in a bilayer Hubbard model at half-filling}

\author{Fangze Liu}
\affiliation{Department of Physics, Stanford University, Stanford, California 94305, USA}
\affiliation{Stanford Institute for Materials and Energy Sciences, SLAC National Accelerator Laboratory, 2575 Sand Hill Road, Menlo Park, California 94025, USA}

\author{Xu-Xin Huang}
\affiliation{Stanford Institute for Materials and Energy Sciences, SLAC National Accelerator Laboratory, 2575 Sand Hill Road, Menlo Park, California 94025, USA}
\affiliation{Department of Applied Physics, Stanford University, Stanford, California 94305, USA}

\author{Edwin W. Huang}
\affiliation{Department of Physics and Institute of Condensed Matter Theory, University of Illinois at Urbana-Champaign, Urbana, Illinois 61801, USA}
\affiliation{Department of Physics and Astronomy, University of Notre Dame, Notre Dame, Indiana 46556, United States}
\affiliation{Stavropoulos Center for Complex Quantum Matter, University of Notre Dame, Notre Dame, Indiana 46556, United States}

\author{Brian Moritz}
\affiliation{Stanford Institute for Materials and Energy Sciences, SLAC National Accelerator Laboratory, 2575 Sand Hill Road, Menlo Park, California 94025, USA}

\author{Thomas P. Devereaux}
\affiliation{Stanford Institute for Materials and Energy Sciences, SLAC National Accelerator Laboratory, 2575 Sand Hill Road, Menlo Park, California 94025, USA}
\affiliation{Department of Materials Science and Engineering, Stanford University, Stanford, California 94305, USA}

\maketitle

In this Supplemental material, we provide additional data and analyses: (\ref{appendix:1}) detailed derivations, measurements, and discussions related to pair-fields; (\ref{appendix:2}) additional measurements on transport properties demonstrating the metal-insulator transition at half-filling; (\ref{appendix:3}) analyses of spin and charge properties in both half-filled and doped scenarios; (\ref{appendix:4}) extensive plots representing the Fermi surface for each layer; (\ref{appendix:5}) a brief introduction of DQMC algorithm.

\section{More Details about Pair-Fields} \label{appendix:1}

\begin{figure*}[htb!] 
\includegraphics[width=0.9\linewidth]{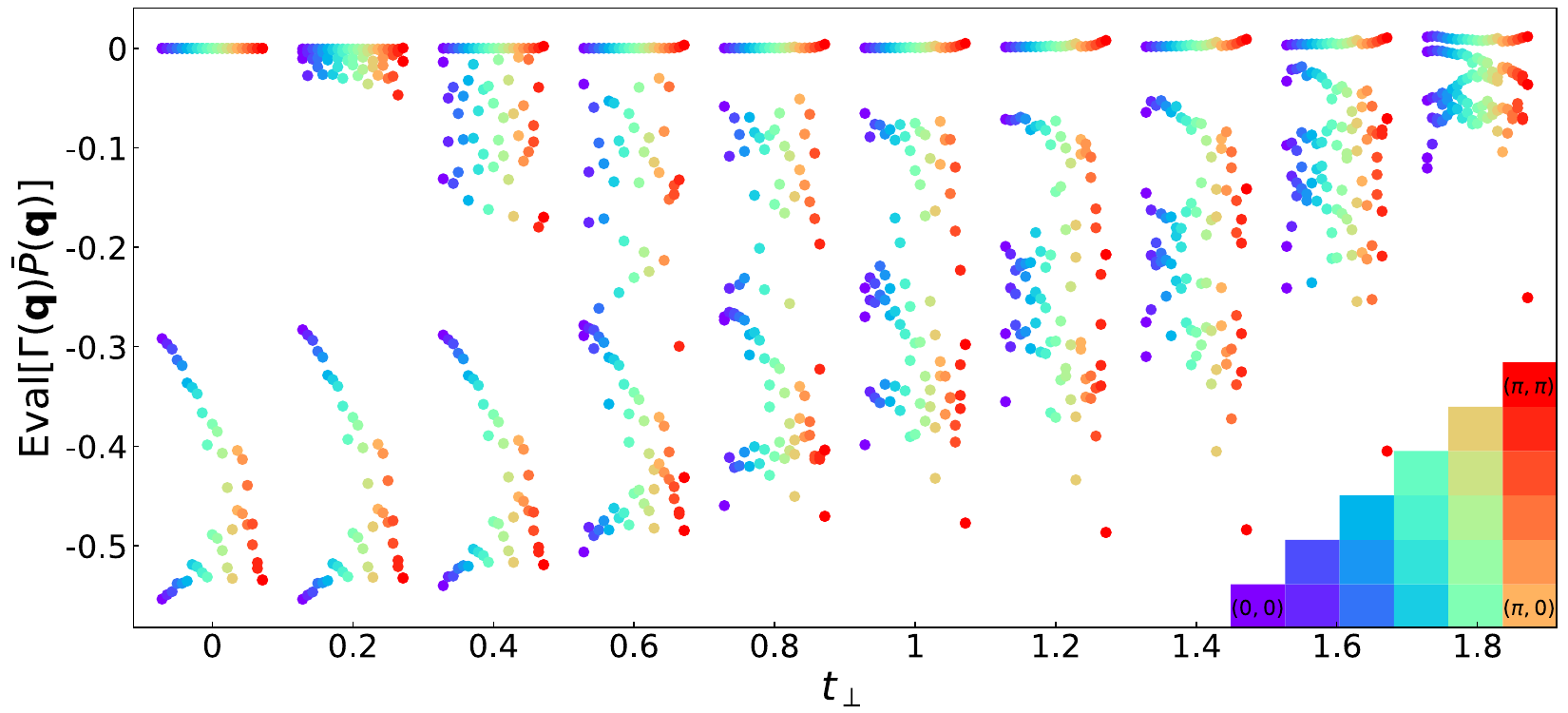}
\caption{\label{pair_evals_halffill}
\textbf{Eigenvalues of $\Gamma(\mathbf{q})\bar{P}(\mathbf{q})$ versus $t_{\perp}$.}
The pair-field matrix eigenvalues are shown for each value of $t_{\perp}$ over a spread of momentum $\mathbf{q}$. Each momentum point in one octant of the Brillouin zone has been assigned a unique color, defined in the color map. Measurements are performed on $10\times 10$ clusters at $\beta=30$ with both layers at half-filling.
}
\end{figure*} 

\begin{figure*}[htb!]  
\includegraphics[width=0.95\linewidth]{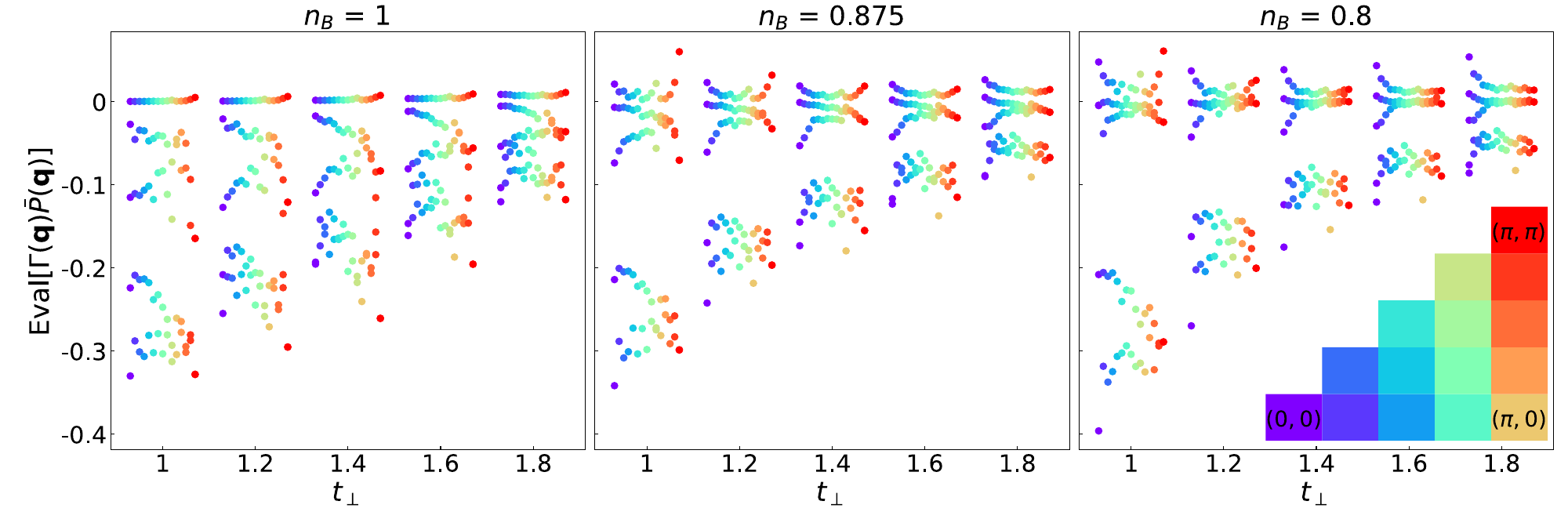}
\caption{\label{pair_evals_dopingB}
\textbf{Comparison between half-filling and hole dopings in layer $B$.} Eigenvalues of $\Gamma(\mathbf{q})\bar{P}(\mathbf{q})$ (colored by center-of-mass momentum $\mathbf{q}$) for $n_B=1, 0.875, 0.8$ in the crossover region as a function of $t_{\perp}$. Measurements are performed on $N=8\times 8$ clusters at $\beta=8$ with $n_A=1\pm 0.004$.
}
\end{figure*}

\begin{figure*}[htb!]  
\includegraphics[width=0.95\linewidth]{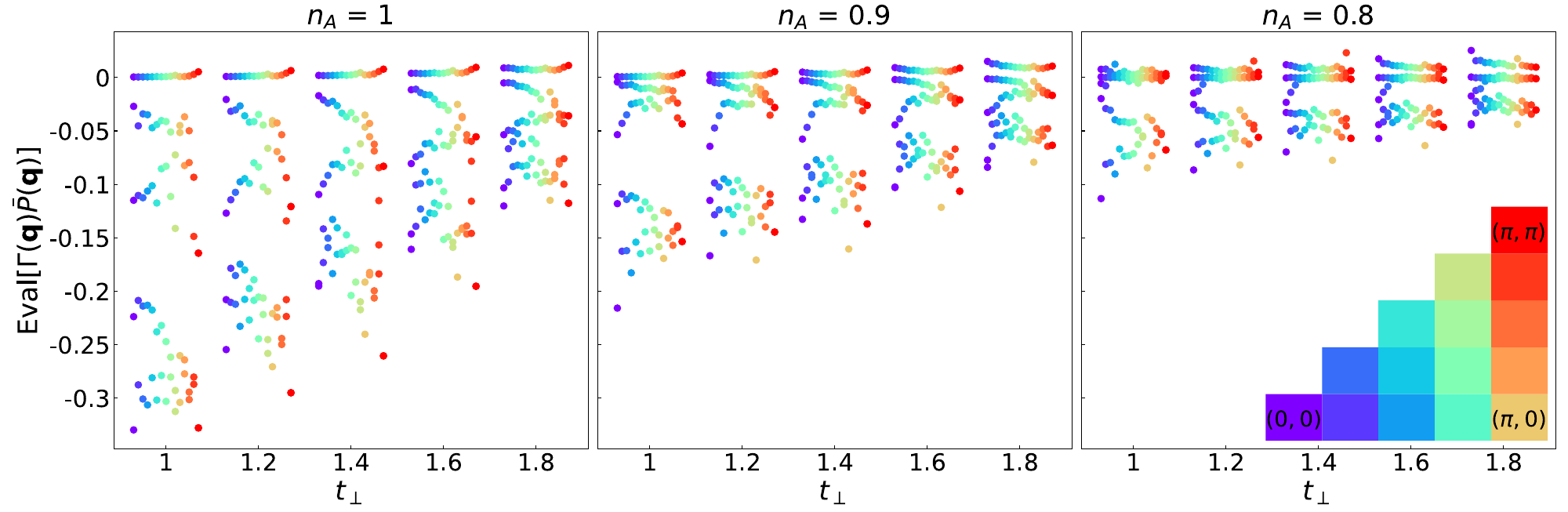}
\caption{\label{pair_evals_dopingA}
\textbf{Comparison between half-filling and hole dopings in layer $A$.} Eigenvalues of $\Gamma(\mathbf{q})\bar{P}(\mathbf{q})$ (colored by center-of-mass momentum $\mathbf{q}$) for $n_A=1, 0.9, 0.8$ in the crossover region as a function of $t_{\perp}$. Measurements are performed on $N=8\times 8$ clusters at $\beta=8$ with $n_B=1\pm 0.03$.
}
\end{figure*}

\textbf{Pair-Field Susceptibilities.}
The full pair-field susceptibility $P(\mathbf{q})$ is given by
\begin{equation}
\begin{aligned}
P_{ab}(\mathbf{q}) = \int_0^{\beta} d\tau \langle \frac{1}{4} \sum_{i, j} e^{i \mathbf{q}\cdot(\mathbf{r}_i-\mathbf{r}_j)} \big(\hat{c}_{i+a \uparrow}(\tau) \hat{c}_{i \downarrow}(\tau)
+ \hat{c}_{i \uparrow}(\tau) \hat{c}_{i+a \downarrow}(\tau)\big) \big(\hat{c}_{j \downarrow}^{\dagger} \hat{c}_{j+b \uparrow}^{\dagger}
+ \hat{c}_{j+b \downarrow}^{\dagger} \hat{c}_{j \uparrow}^{\dagger}\big)
\rangle,
\end{aligned}
\end{equation}
and the uncorrelated pair-field susceptibility $\bar{P}(\mathbf{q})$ is given by
\begin{equation}
\begin{aligned}
\bar{P}_{ab}(\mathbf{q}) = \int_0^{\beta} d\tau \frac{1}{4} \sum_{i, j} e^{i \mathbf{q}\cdot(\mathbf{r}_i-\mathbf{r}_j)}
\Big(
&\langle \hat{c}_{i+a \uparrow}(\tau) \hat{c}_{j+b \uparrow}^{\dagger} \rangle
  \langle \hat{c}_{i \downarrow}(\tau) \hat{c}_{j \downarrow}^{\dagger} \rangle +\langle \hat{c}_{i+a \uparrow}(\tau) \hat{c}_{j \uparrow}^{\dagger} \rangle
  \langle \hat{c}_{i \downarrow}(\tau) \hat{c}_{j+b \downarrow}^{\dagger} \rangle \\
&+\langle \hat{c}_{i \uparrow}(\tau) \hat{c}_{j+b \uparrow}^{\dagger} \rangle
  \langle \hat{c}_{i+a \downarrow}(\tau) \hat{c}_{j \downarrow}^{\dagger} \rangle 
 +\langle \hat{c}_{i \uparrow}(\tau) \hat{c}_{j \uparrow}^{\dagger} \rangle
  \langle \hat{c}_{i+a \downarrow}(\tau) \hat{c}_{j+b \downarrow}^{\dagger} \rangle
\Big),
\end{aligned}
\end{equation}
where layer indices have been suppressed for clarity.\\

\textbf{Eigenvalues of $\Gamma(\mathbf{q})\bar{P}(\mathbf{q})$.}
In the main text, we show only the lowest eigenvalues of $\Gamma(\mathbf{q})\bar{P}(\mathbf{q})$ at $\mathbf{q}=(0,0)$ and $(\pi,\pi)$. In Fig.~\ref{pair_evals_halffill}, we show all the eigenvalues for $\Gamma(\mathbf{q})\bar{P}(\mathbf{q})$ at different momenta $\mathbf{q}$. The pair-field susceptibilities have both inversion and C4-rotational symmetry, so we show values just in one octant. We consider only nearest-neighbor pairing giving 5 eigenvalues for each $\mathbf{q}$. The largest negative eigenvalue for $\mathbf{q}=(\pi,\pi)$ has an additional two-fold degeneracy, since eigenstates $\mathbf{v}=(a,-a,-b,b,0)$ and $\mathbf{v}=(a,a,-b,-b,0)$ are equivalent under C4 symmetry ($a, b \in \mathbb{C}$). 
For $t_{\perp} \sim \sqrt{2}$, the eigenvalue with $\mathbf{q}=(\pi,\pi)$ clearly is the smallest among all $\mathbf{q}$. 

A comparison to the results for hole dopings are shown in Fig.~\ref{pair_evals_dopingB} and Fig.~\ref{pair_evals_dopingA}. Interestingly, when layer $B$ is $1/8$ hole doped or layer $A$ is hole doped up to $1/5$, the largest negative eigenvalue resides at $\mathbf{q}=(\pi,0)$ with $\mathbf{v}=(1,0,0,0,0)$, being marginally smaller than those of other $\mathbf{q}$ in the $t_{\perp}\sim \sqrt{2}$ regime.

\begin{figure*}[htb!]  
\includegraphics[width=0.99\linewidth]{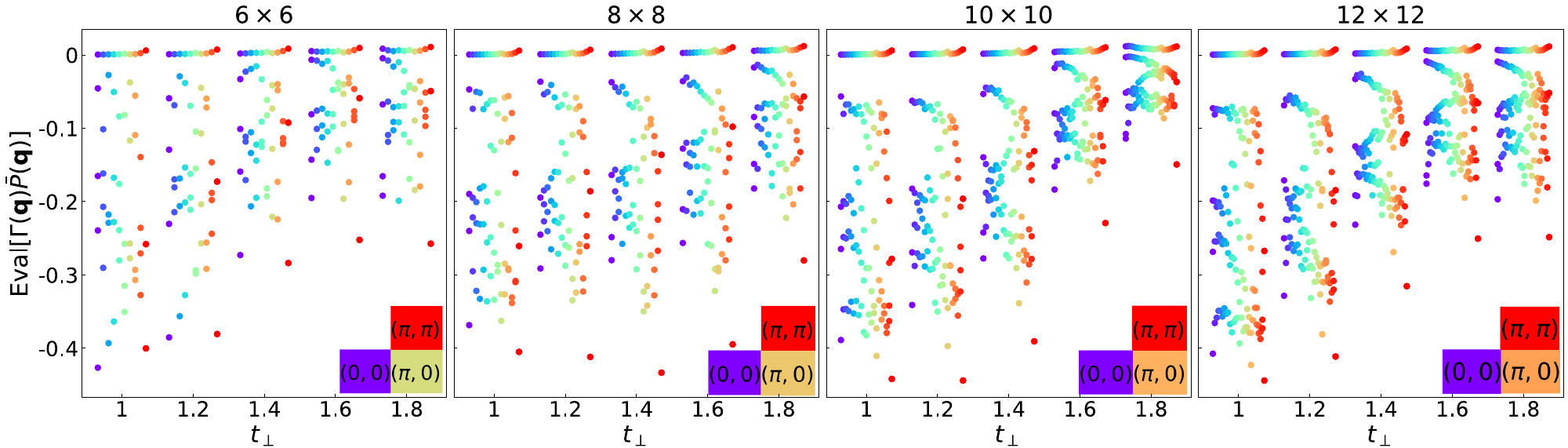}
\caption{\label{pair_evals_finitesize}
\textbf{Eigenvalues of $\Gamma(\mathbf{q})\bar{P}(\mathbf{q})$ for various cluster sizes.} Eigenvalues of $\Gamma(\mathbf{q})\bar{P}(\mathbf{q})$ (colored by center-of-mass momentum $\mathbf{q}$) for different cluster sizes. Measurements are performed at $\beta=12$ with both layers at half-filling.
}
\end{figure*}

We also investigated finite-size effects in Fig.~\ref{pair_evals_finitesize} by considering different clusters up to size $12\times 12$. There is a non-negligible finite-size effect on the eigenvalues of $\Gamma(\mathbf{q})\bar{P}(\mathbf{q})$; however, the dominant eigenvalue for $J_{\text{K}} \sim J_{\text{H}}$ occurs at $\mathbf{q}=(\pi,\pi)$ for all cluster sizes.\\

\begin{figure}[b!] 
\includegraphics[width=0.9\linewidth]{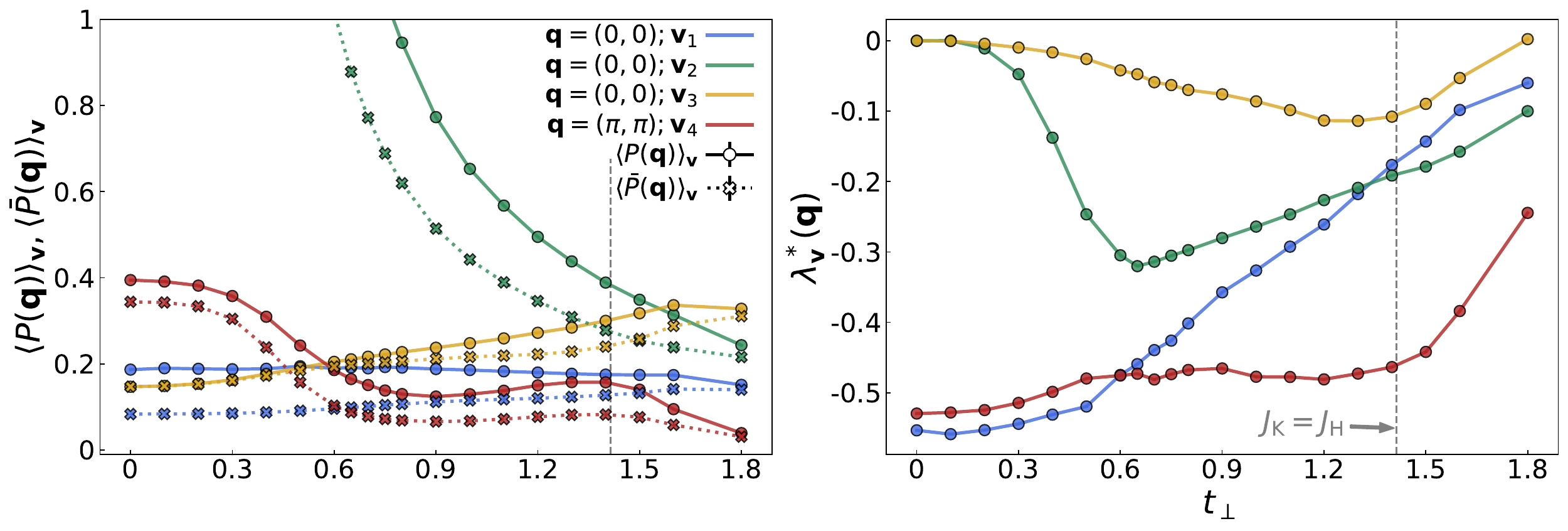}
\caption{\label{pair_vec_appendix}
\textbf{Pair fields for trial vectors.}
Top panel: $\langle P(\mathbf{q})\rangle_{\mathbf{v}}$ (solid lines) and $\langle \bar{P}(\mathbf{q})\rangle_{\mathbf{v}}$ (dashed lines). Bottom panel: $\langle \Gamma(\mathbf{q}) \bar{P}(\mathbf{q})\rangle_{\mathbf{v}}$ for different center-of-mass momenta $\mathbf{q}$. Trial vectors $\mathbf{v}$: $\mathbf{v}_1=(1,-1,0,0,0)/\sqrt{2}$, $\mathbf{v}_2=(0,0,1,-1,0)/\sqrt{2}$, $\mathbf{v}_3=(0,0,0,0,1)$, $\mathbf{v}_4=(1,-1,-1,1,0)/2$. Measurements performed on $N=10\times10$ clusters at $\beta=30$.}
\end{figure}

\textbf{Pair-Field Measurements for Trial Vectors.}
\label{appendix:pair}
To better understand the pairing structure, we estimate the pairing strength for some trial vectors that correspond closely to expected eigenvectors for different values of $t_{\perp}$. The operator for a mode $\mathbf{\alpha}$ as a function of the center-of-mass momentum $\mathbf{q}$ is commonly expressed as
\begin{equation}\label{eq:Delta_a}
\begin{aligned}
\hat{\Delta}_{\alpha, \ell} (\mathbf{q}) 
&= \sum_{\mathbf{k}} f_{\alpha}(\mathbf{k})  \hat{c}_{(\mathbf{k}+\mathbf{q}) \ell \uparrow} \hat{c}_{(-\mathbf{k}) \ell \downarrow}.
\end{aligned}
\end{equation}
For example, the operators in the d$_{x^2-y^2}$-wave channel, with form factor $f_d(\mathbf{k}) = \cos{k_x}-\cos{k_y}$, and the extended s-wave channel, with $f_{s*}(\mathbf{k}) = \cos{k_x}+\cos{k_y}$, are 
\begin{equation}
\begin{aligned}
\hat{\Delta}_{d,\ell} (\mathbf{q})
&= \sum_i e^{i \mathbf{q}\cdot \mathbf{r}_i} (\hat{\varsigma}_{i,x,\ell} - \hat{\varsigma}_{i,y,\ell}), \\
\hat{\Delta}_{s*,\ell} (\mathbf{q})
&= \sum_{i} e^{i \mathbf{q}\cdot \mathbf{r}_i} (\hat{\varsigma}_{i,x,\ell} + \hat{\varsigma}_{i,y,\ell}), \\
\end{aligned}
\end{equation}
respectively, where $\hat{\varsigma}_{i,a, \ell} =(\hat{c}_{(i+a) \ell \uparrow}\hat{c}_{i \ell \downarrow}+\hat{c}_{i \ell \uparrow}\hat{c}_{(i+a) \ell \downarrow})/2$ is the singlet pair operator on the bond from $i$ to $i+a$ in layer $\ell$. Then, the correlated pair-field susceptibility can be defined as 
\begin{equation} \label{eq: P_sym}
\langle P (\mathbf{q}) \rangle_{\alpha, \ell} = \int_0^{\beta} d\tau \langle \hat{\Delta}_{\alpha, \ell} (\tau, \mathbf{q}) \hat{\Delta}_{\alpha, \ell}^{\dagger} (\mathbf{q}) \rangle,
\end{equation}
and its disconnected part is denoted as $\langle \bar{P}(\mathbf{q}) \rangle_{\alpha, \ell}$. 

Consider an arbitrary vector $\mathbf{v}$ in the basis $(\hat{\varsigma}_{\mathbf{q},x,A}, \hat{\varsigma}_{\mathbf{q},y,A}, \hat{\varsigma}_{\mathbf{q},x,B}, \hat{\varsigma}_{\mathbf{q},y,B}, \hat{\varsigma}_{\mathbf{q},z})$, representing the form of a singlet pair, as in the main text. For example, $\Delta_{d,A}$, as defined in Eqn.~\ref{eq:Delta_a}, implies $\mathbf{v} = (1,-1,0,0,0)/\sqrt{2}$; $\mathbf{v}=(1,-1,-1,1,0)/2$ at $\mathbf{q}=(\pi,\pi)$ corresponds to a PDW with $(\hat{\varsigma}_{\mathbf{q},x,A}-\hat{\varsigma}_{\mathbf{q},y,A}-\hat{\varsigma}_{\mathbf{q},x,B}+\hat{\varsigma}_{\mathbf{q},y,B})$ at $\mathbf{q}= (\pi,\pi)$. 
Therefore, we can express the correlated pair-field susceptibility for an arbitrary vector $\mathbf{v}$ as
\begin{equation} \label{eq: P_vec}
\langle P(\mathbf{q})\rangle_{\mathbf{v}} = \bra{\mathbf{v}} P(\mathbf{q}) \ket{\mathbf{v}}.
\end{equation}


The nonsymmetric diagonalizable matrix $\Gamma(\mathbf{q})\bar{P}(\mathbf{q})$, which has eigenvalues $\lambda_i(\mathbf{q})$, right eigenvectors $\mathbf{\tilde{\phi}}_i(\mathbf{q})$ and left eigenvectors $\mathbf{\phi}_i(\mathbf{q})$, can be expressed as
\begin{equation} \label{eq:Vchib1}
\Gamma \bar{P}
= \sum_i \ket*{\tilde{\mathbf{\phi}}_i} \lambda_i \bra{\mathbf{\phi}_i},
\end{equation}
where the $\mathbf{q}$-dependence is implied. 
For an invertible Hermitian matrix $\bar{P}$ we have $\bar{P}\bar{P}^{-1}=I$. Then,
\begin{equation}
\begin{aligned}
& \bar{P} \Gamma \ket{\mathbf{\phi}_i} = \lambda_i \ket{\mathbf{\phi}_i}, \\
& \Gamma \ket{\mathbf{\phi}_i} = \lambda_i \bar{P}^{-1 }\ket{\mathbf{\phi}_i}, \\
&  \Gamma \bar{P} \big( \bar{P}^{-1} \ket{\mathbf{\phi}_i} \big) =  \lambda_i \big( \bar{P}^{-1 }\ket{\mathbf{\phi}_i} \big). \\
\end{aligned}
\end{equation}

With $\Gamma \bar{P} \ket*{\tilde{\mathbf{\phi}}_i} = \lambda_i \ket*{\tilde{\mathbf{\phi}}_i}$ and the normalization of the inner product of left and right eigenvectors $\bra*{\mathbf{\phi}_i}\ket{\tilde{\mathbf{\phi}}_j}=\delta_{ij}$, the left and right eigenvectors are related by
\begin{equation}
\ket{\mathbf{\phi}_i} = \mathcal{N}_{i} \bar{P} \ket*{\tilde{\mathbf{\phi}}_i} \text{, \;\;\;\;} \mathcal{N}_{i}=1/\bra{\tilde{\phi}_i}\bar{P} \ket{\tilde{\phi}_i}. 
\end{equation}

The full pair-field susceptibility can be written in terms of these eigenvectors by substituting $\Gamma \bar{P}$ such that
\begin{equation}
\begin{aligned}
P &= \bar{P} \left(\mathbb{1} + \Gamma \bar{P}\right)^{-1} 
= \sum_{i} \bar{P} \ket*{\tilde{\mathbf{\phi}}_i} (1+\lambda_i)^{-1} \bra{\mathbf{\phi}_i}\\
&= \sum_{i} \frac{1}{\mathcal{N}_{i}} \ket{\mathbf{\phi}_i} (1+\lambda_i)^{-1} \bra{\mathbf{\phi}_i}, \\
\end{aligned}
\end{equation}
so the vector $\ket{\mathbf{\phi}_i}$ having $\lambda_i \rightarrow -1$ is then the leading eigenvector of $P$.

To better understand the tendencies for trial vectors $\mathbf{v}$, they can be decomposed in the eigenbasis of $\Gamma\bar{P}$, such that $\mathbf{v} = \sum_i c_i \mathbf{\phi}_i$. A variational estimator, defined as
\begin{equation} \label{eq:Vchib2}
\begin{aligned}
\lambda^*_{\mathbf{v}}
& = \frac{\bra{\mathbf{v}} \Gamma \ket{\mathbf{v}}}{\bra{\mathbf{v}} \bar{P}^{-1} \ket{\mathbf{v}}} 
= \frac{\sum_{ij} c_i^* c_j \mathcal{N}_{j} \bra{\mathbf{\phi}_i} \Gamma \bar{P} \ket*{\tilde{\mathbf{\phi}}_j}}{\sum_{ij} c_i^* c_j \mathcal{N}_{j} \bra{\mathbf{\phi}_i} \bar{P}^{-1} \bar{P} \ket*{\tilde{\mathbf{\phi}}_j}} \\
&= \frac{\sum_{ijk} c_i^* c_j \mathcal{N}_{j} \braket{\mathbf{\phi}_i}{\tilde{\mathbf{\phi}_k}} \lambda_k \braket{\mathbf{\phi}_k}{\tilde{\mathbf{\phi}_j}}}{\sum_{i} |c_i|^2 \mathcal{N}_{i}} \\
&=\frac{\sum_{i} |c_i|^2 \mathcal{N}_{i} \; \lambda_i}{\sum_{i} |c_i|^2 \mathcal{N}_{i}},
\end{aligned}
\end{equation}
provides a direct measure as combinations of the eigenvalues  $\lambda_i(\mathbf{q})$ of $\Gamma(\mathbf{q}) \bar{P}(\mathbf{q})$ weighted by $|c_i(\mathbf{q})|^2\mathcal{N}_{i}(\mathbf{q})/ \big(\sum_{i} |c_i(\mathbf{q})|^2 \mathcal{N}_{i}(\mathbf{q})\big)$.\\

Figure~\ref{pair_vec_appendix} illustrates $\langle P(\mathbf{q}) \rangle_{\mathbf{v}}$, $\langle \bar{P}(\mathbf{q}) \rangle_{\mathbf{v}}$, and $\lambda^*_{\mathbf{v}}$, for selected trial vectors as a function of $t_{\perp}$. 
For $\mathbf{q}=(0,0)$, the chosen vectors are: 
(1) $\mathbf{v}=(1,-1,0,0,0)/\sqrt{2}$, characterizing the leading eigenvector for $0\leq t_{\perp}<0.7$; 
(2) $\mathbf{v}=(0,0,1,-1,0)/\sqrt{2}$, representing the uniform $d$-wave pairing in layer $B$, which surpasses the uniform $d$-wave pairing in layer $A$ when $t_{\perp}>\sqrt{2}$; 
(3) $\mathbf{v}=(0,0,0,0,1)$, signifying the interlayer pairing, which is enhanced as $J_{\text{K}} \sim J_{\text{H}}$. It is noteworthy that this uniform interlayer pair-field susceptibility outperforms in the large $t_{\perp}$ regime, while its pairing interaction peaks at $t_{\perp}=\sqrt{2}$. 
For $\mathbf{q}=(\pi,\pi)$, we focus on $\mathbf{v}=(1,-1,-1,1,0)/2$, a representative vector in the putative PDW region, where $\langle P(\pi,\pi) \rangle_{\mathbf{v}}$ and $|\lambda^*_{\mathbf{v}}(\pi,\pi)|$ are notably enhanced at $t_{\perp}=\sqrt{2}$.

\section{Metal-Insulator Transition at Half-Filling} \label{appendix:2} 

\begin{figure}
    \centering
    \includegraphics[width=0.45\linewidth]{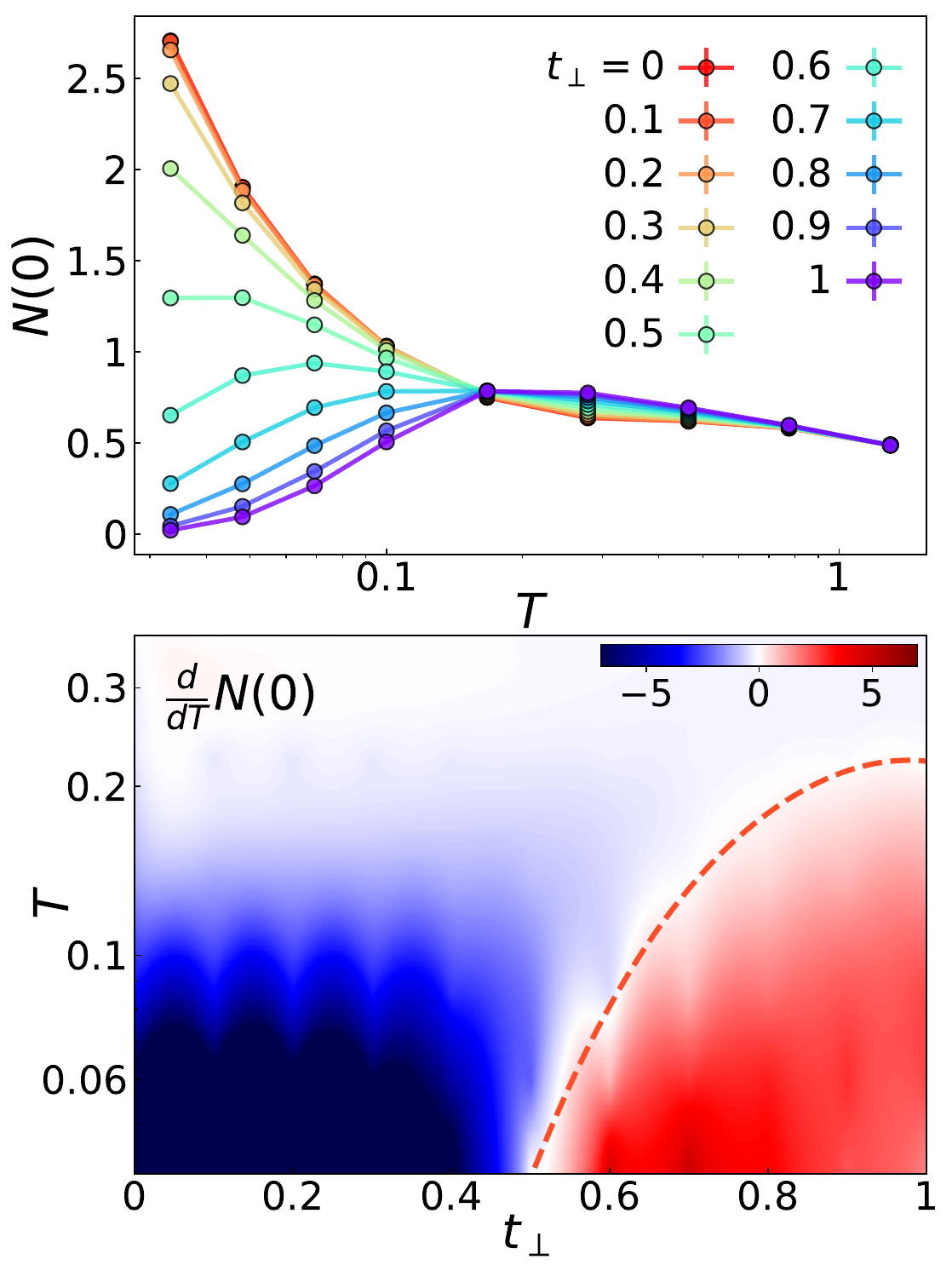}
    \includegraphics[width=0.45\linewidth]{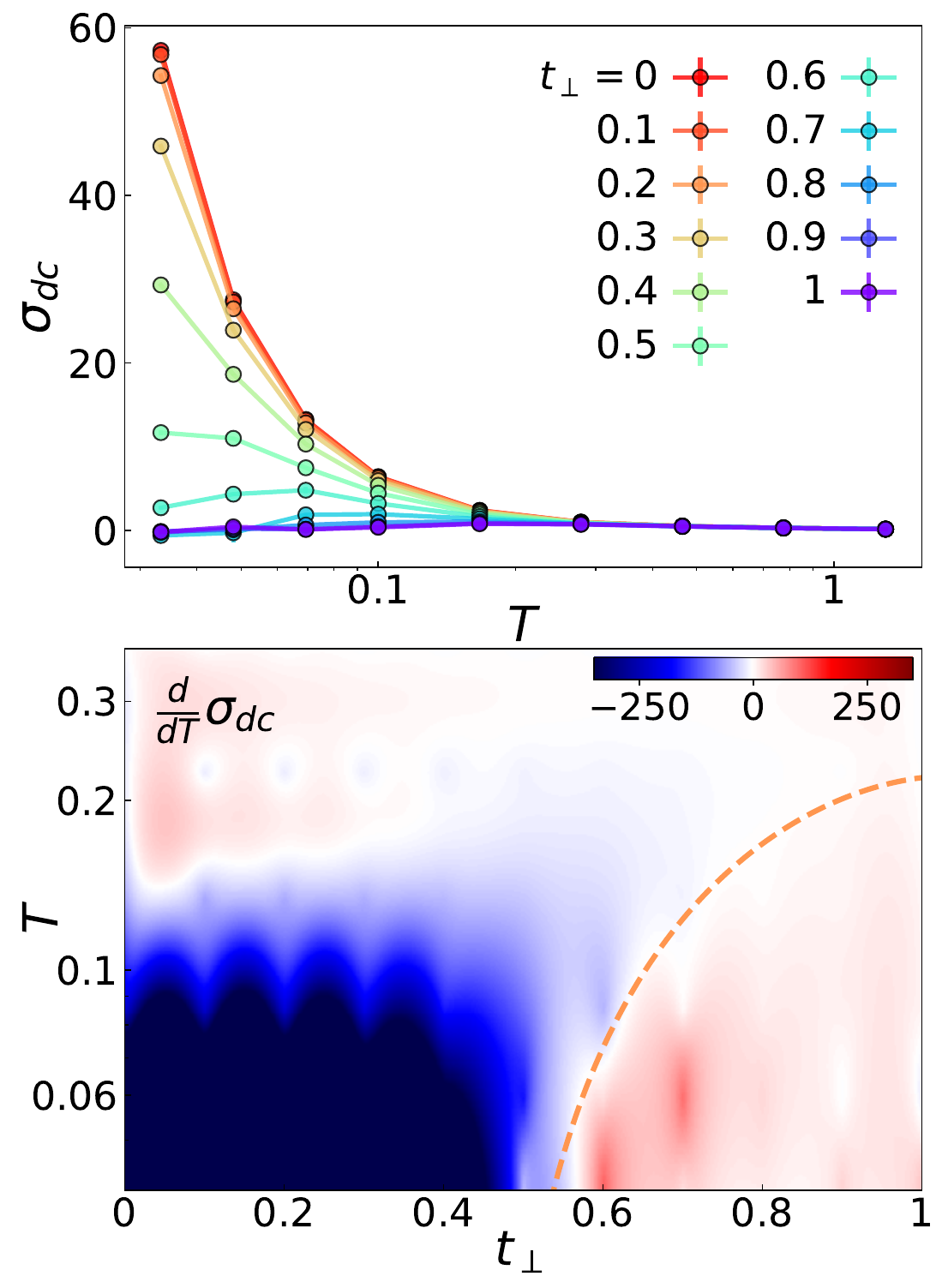}\\
    \caption{\label{transport}
    \textbf{Transport properties.}
    \textit{Left}: Density of states at the Fermi energy, evaluated by the proxy $N(\omega=0)=\sum_{\mathbf{k}}\beta G(\mathbf{k},\tau=\frac{\beta}{2})/\pi N$. 
    \textit{Right}: DC conductivity $\sigma_{dc}$. 
    \textit{Top}: Both $\mathbf{N(0)}$ and $\mathbf{\sigma_{dc}}$ are plotted against temperatures for various values of $t_{\perp}$. An increase in $N(0)$ and $\sigma_{dc}$ with decreasing $T$ signifies metallic behavior, while a decreasing trend indicates insulating behavior. 
    \textit{Bottom}: Temperature derivative of $\mathbf{N(0)}$ and $\mathbf{\sigma_{dc}}$. Dashed lines mark the transition between metallic and insulating behaviors. Temperature axes are displayed on a log scale. All results are obtained on $10\times 10$ lattices. }
\end{figure}

To distinguish between metallic and insulating behavior, we can use the density of states, through the single-particle spectral function $A(\omega)$. The imaginary time Green's function $G(\mathbf{k},\tau) = \langle c_{\mathbf{k}}(\tau) c^{\dagger}_{\mathbf{k}} \rangle$ by
\begin{equation}
\begin{aligned}
G(\mathbf{k}, \tau) 
&= \int_{-\infty}^{\infty} d\omega \frac{e^{-\omega \tau}}{1+e^{-\beta \omega}} A(\mathbf{k}, \omega)
= \int_{-\infty}^{\infty} d\omega \frac{e^{-\omega (\tau-\beta/2)}}{2 \cosh{(\beta \omega/2)}} A(\mathbf{k}, \omega),
\end{aligned}
\end{equation}
and analytic continuation can be used to determine the single-particle density of states $N(\omega) = \sum_{\mathbf{k}} A(\mathbf{k}, \omega)/N$ from the imaginary-time QMC data.
To avoid numerical complications associated with analytic continuation, we can estimate
\begin{equation}
N(0) \approx \frac{\beta}{\pi N} \sum_{\mathbf{k}} G\left(\mathbf{k}, \tau=\frac{\beta}{2}\right).
\end{equation}

Figure~\ref{transport} identifies the location of the metal-insulator transition boundary (dashed line). The temperature behavior of the single-particle density of states at the Fermi energy $N(0)$ indicates a metallic phase in the blue shaded region at low $t_{\perp}$, where $N(0)$ increases as $T$ decreases, and an insulating phase in the red shaded region at high $t_{\perp}$, where $N(0)$ decreases as $T$ decreases.

The metal-insulator transition also can be characterized by the dc conductivity $\sigma_{dc}$, which is related to the current-current correlation function $\Lambda$. The imaginary time current-current correlator $\mathbf{\Lambda}(\tau)$ at momentum zero can be obtained through DQMC simulations by measuring
\begin{equation}
\mathbf{\Lambda}(\tau) = \langle \hat{\mathbf{j}}(\tau) \hat{\mathbf{j}} \rangle,
\end{equation}
where
\begin{equation}
\hat{\mathbf{j}} = i \sum_{i j \sigma} t_{i j} (\mathbf{r}_i - \mathbf{r}_j) \hat{c}^{\dagger}_{i \sigma} \hat{c}_{j \sigma}.
\end{equation}

To obtain the dc conductivity from the imaginary time response, we can either perform analytic continuation to compute $\text{Im}\mathbf{\Lambda}(\omega)$ for all frequencies, which is related to $\mathbf{\Lambda}(\tau)$ through the following relation:
\begin{equation}
\mathbf{\Lambda}(\tau) = \int_{-\infty}^{\infty} \frac{d\omega}{\pi} \frac{e^{-\omega \tau}}{1-e^{-\beta \omega}} \text{Im}\mathbf{\Lambda}(\omega).
\end{equation}
or estimate the low frequency behavior of $\text{Im}\mathbf{\Lambda} \approx \omega \mathbf{\sigma}_{dc}$ at sufficient low temperatures from
\begin{equation}
\Lambda_{xx}\left(\tau=\frac{\beta}{2}\right) \approx \frac{\pi \sigma_{dc}}{\beta^2}.
\end{equation}

Figure~\ref{transport} captures the metal-insulator transition boundary from the conductivity $\sigma_{dc}$, which decreases as $T$ decreases indicating insulating behavior, or increasing as $T$ decreases indicating metallic behavior. This generally matches the boundary obtained from the density of states at the Fermi energy. At $\beta=30$, the metal-insulator transition occurs at $t_{\perp} \sim 0.5$, which coincides with the peak in AFM correlations and is concommitant to the crossover from uniform pairing to PDW in Fig.~2 of the main text.

\begin{figure*}[thb] 
\centering
\begin{minipage}{0.85\textwidth}
    \includegraphics[width=\linewidth]{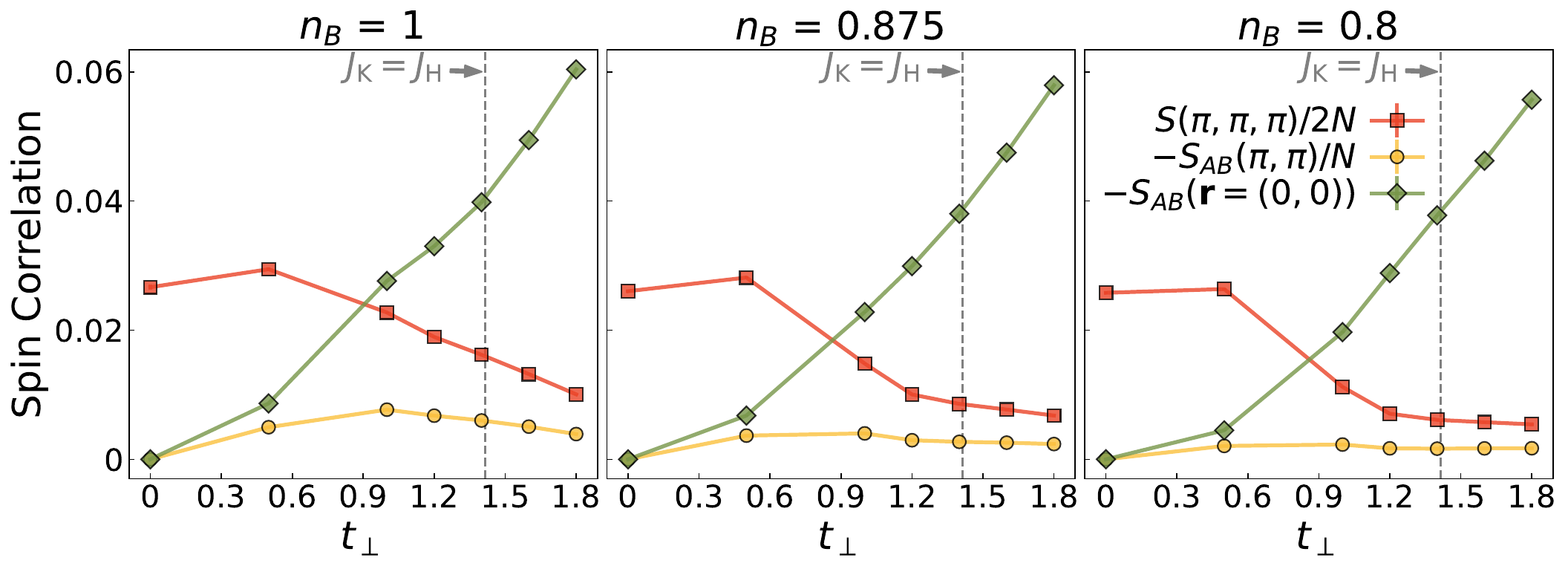}
    \label{ss_dopingB}
\end{minipage}\\
\begin{minipage}{0.85\textwidth}
    \includegraphics[width=\linewidth]{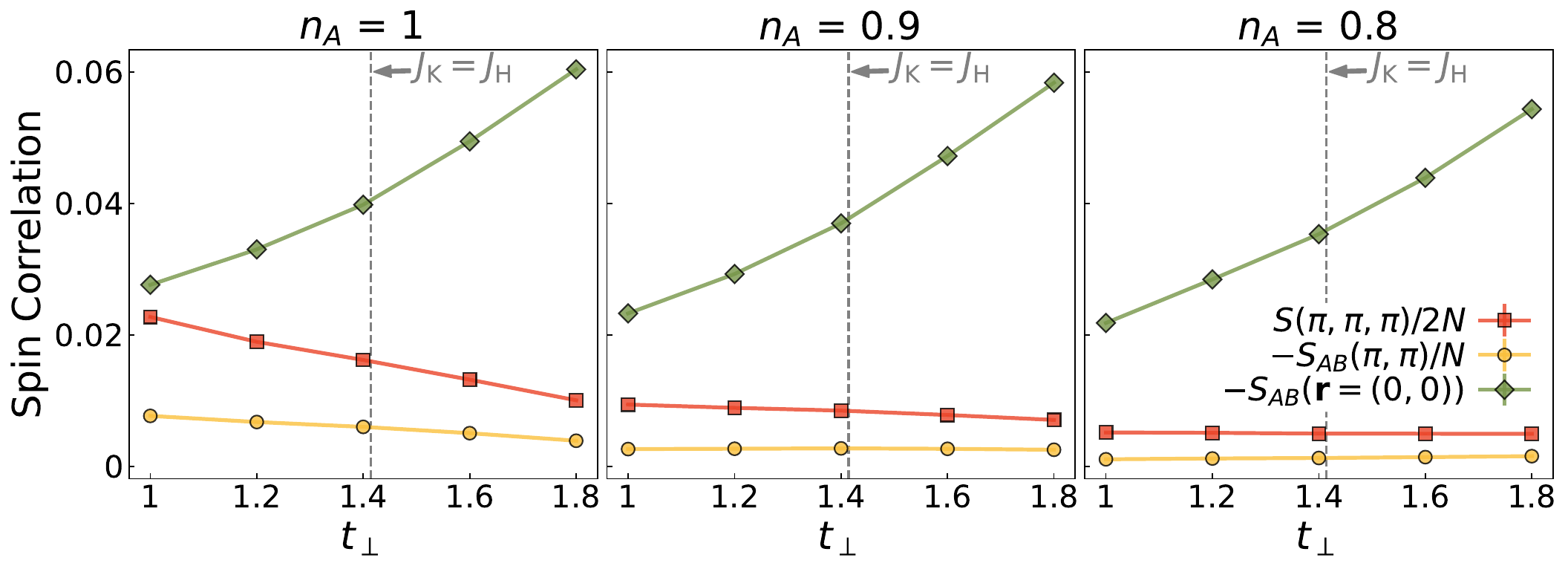}
    \label{ss_dopingA}
\end{minipage}\\
\caption{
\textbf{Equal-time spin-spin correlations} for various hole dopings in (top) layer $B$ and (bottom) layer $A$. This figure is similar to Fig.~1 in the main text, except the common parameters for all plots are $N=8\times 8$, $\beta=8$, and (top) $n_A=1 \pm 0.004$ and (bottom) $n_B=1\pm 0.03$. To mitigate the severe sign problem at $\beta=8$ when layer $A$ is doped and $t_{\perp} < 1$, we restrict the range of $t_{\perp}$ to larger values.}
\label{ss_doping}
\end{figure*}

\begin{figure*}[thb] 
\centering
\begin{minipage}{1\textwidth}
    \includegraphics[width=0.85\linewidth]{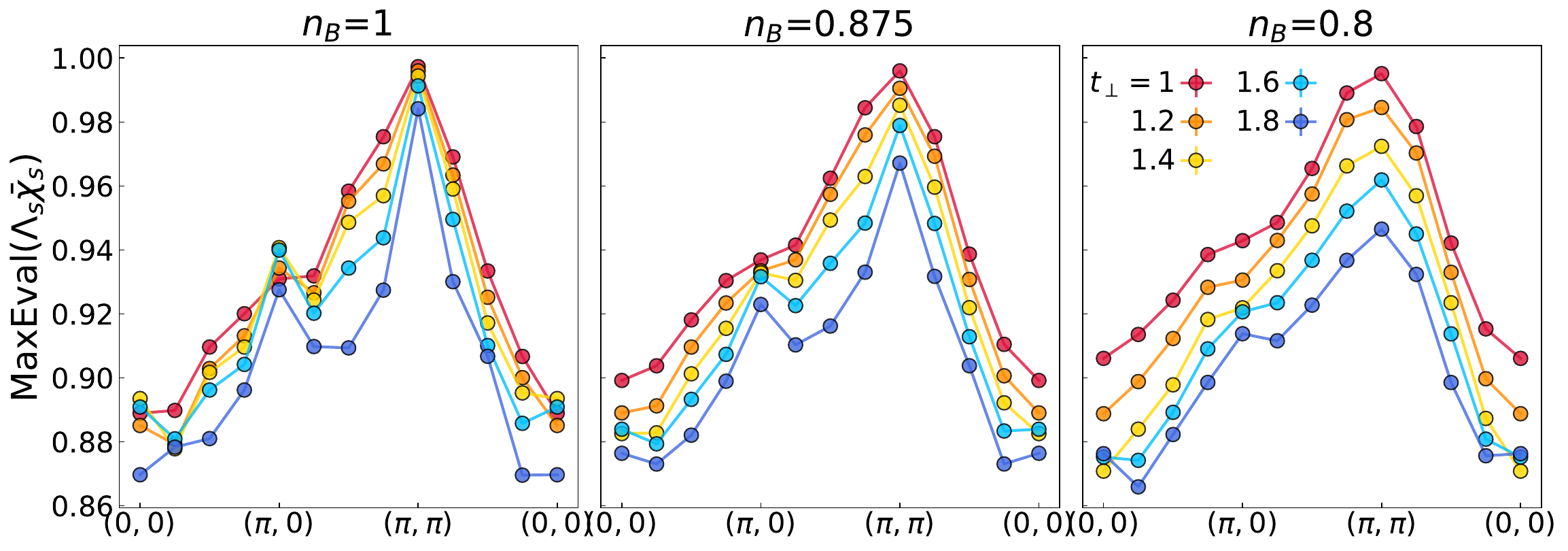}
    \label{spin_maxeval_dopingB}
\end{minipage}\\
\begin{minipage}{1\textwidth}
    \includegraphics[width=0.85\linewidth]{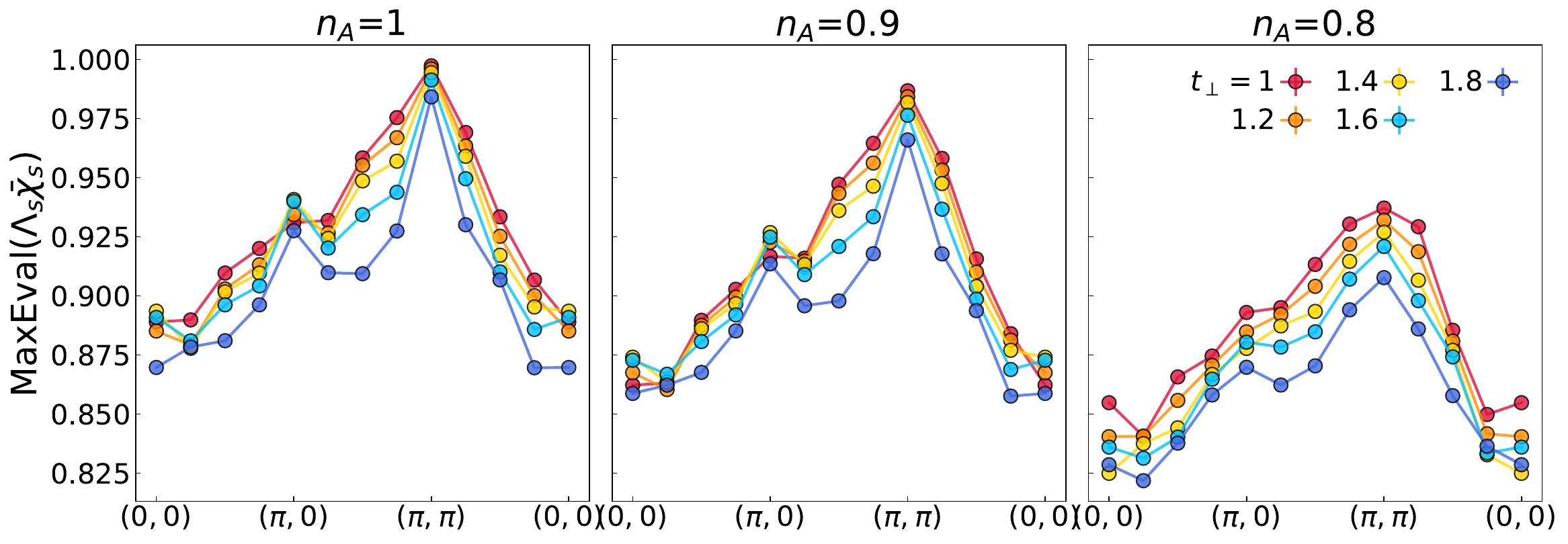}
    \label{spin_maxeval_dopingA}
\end{minipage}\\
\caption{
\textbf{The maximum eigenvalue of $\Lambda_s(\mathbf{q}) \chi_s(\mathbf{q})$} for various hole dopings in (top) layer $B$ and (bottom) layer $A$. The common parameters for all plots are $N=8\times 8$, $\beta=8$, and (top) $n_A=1\pm 0.004$ and (bottom) $n_B=1\pm 0.03$. }
\label{spin_maxeval_doping}
\end{figure*}

\begin{figure*}[thb] 
\centering
\begin{minipage}{1\textwidth}
    \includegraphics[width=0.85\linewidth]{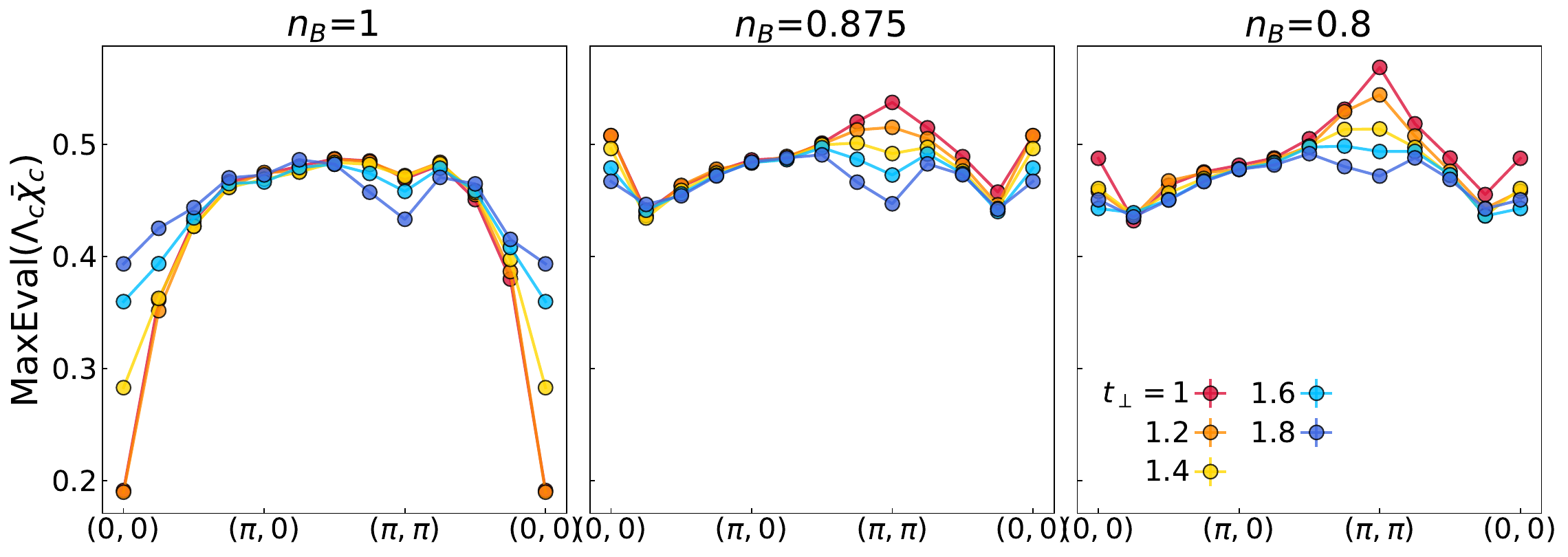}
    \label{charge_maxeval_dopingB}
\end{minipage}\\
\begin{minipage}{1\textwidth}
    \includegraphics[width=0.85\linewidth]{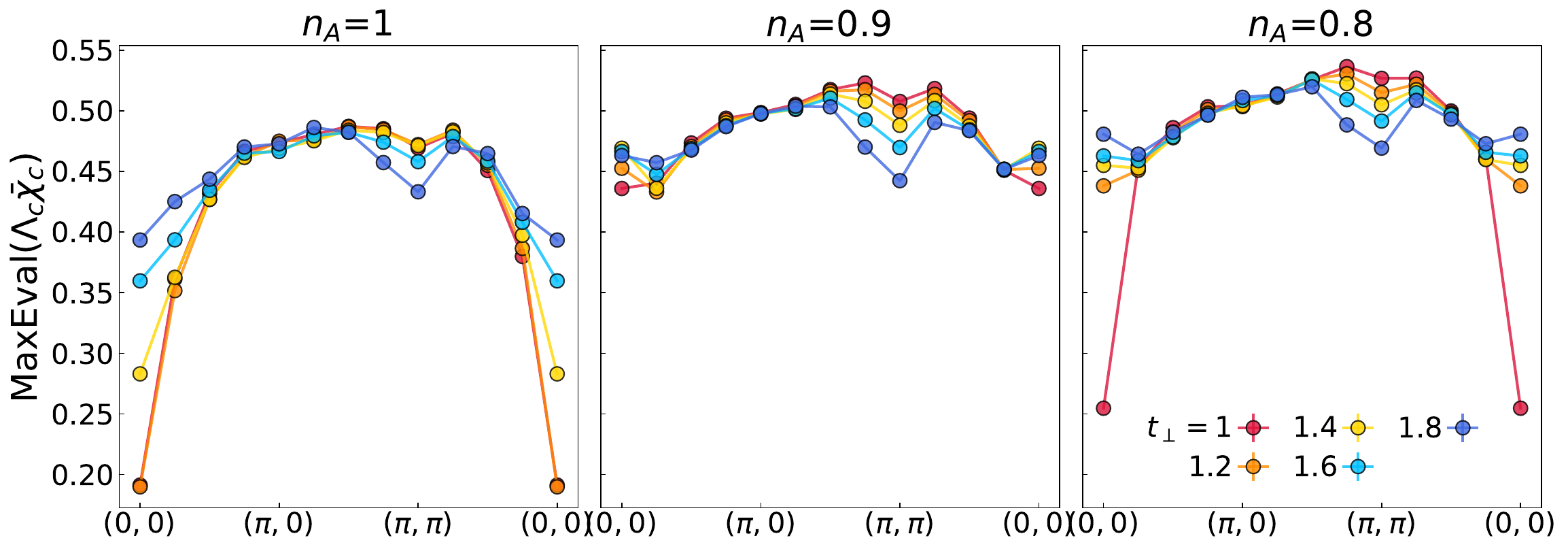}
    \label{charge_maxeval_dopingA}
\end{minipage}\\
\caption{
\textbf{The maximum eigenvalue of $\Lambda_c(\mathbf{q}) \chi_c(\mathbf{q})$} for various hole dopings in (top) layer $B$ and (bottom) layer $A$. The common parameters for all plots are $N=8\times 8$, $\beta=8$, and (top) $n_A=1\pm 0.004$ and (bottom) $n_B=1\pm 0.03$. }
\label{charge_maxeval_doping}
\end{figure*}

\section{Spin and Charge in Doped Systems} \label{appendix:3}

Figure~\ref{ss_doping} shows the variation of the equal-time spin correlations at the antiferromagnetic wavevector $S(\pi,\pi,\pi)$, the interlayer spin singlet correlations $S_{AB}(\mathbf{r}=(0,0))$ and $S_{AB}(\pi,\pi)$ with hole doping in layer $A$ or $B$ (keeping the other layer fixed at half-filling). It shows that doping layer $B$ slightly suppresses the AFM correlations while doping layer $A$ destroys them. 

Since the particle-particle interaction vertex is introduced for pairing, the particle-hole interaction vertex can also be calculated for magnetic susceptibilities. Here we introduce the magnetic susceptibility tensor
\begin{equation}
\begin{aligned}
\chi_{\sigma \sigma^{'}}^{\ell\ell'} (\mathbf{q})
& = \int d\tau \frac{1}{N}\sum_{i,j} e^{i\mathbf{q}\cdot(\mathbf{r_i}-\mathbf{r_j})} \langle \hat{n}_{i \ell \sigma}(\tau) \hat{n}_{i \ell' \sigma'}(0) \rangle,
\end{aligned}
\end{equation}
and the bare magnetic susceptibility tensor
\begin{equation}
\begin{aligned}
\bar{\chi}_{\sigma \sigma^{'}}^{\ell\ell'} (\mathbf{q}) 
&= \delta_{\sigma,\sigma'} \int d\tau \frac{1}{N}\sum_{i,j} e^{i\mathbf{q}\cdot(\mathbf{r_i}-\mathbf{r_j})} \langle \hat{c}^{\dagger}_{i \ell \sigma}(\tau) \hat{c}_{j \ell' \sigma'} \rangle \langle \hat{c}_{i \ell \sigma}(\tau) \hat{c}^{\dagger}_{j \ell' \sigma'} \rangle \\
&= \delta_{\sigma \sigma'} \int d\tau \frac{1}{N}\sum_{i,j} e^{i\mathbf{q}\cdot(\mathbf{r_i}-\mathbf{r_j})} G_{j i; \sigma}^{\ell' \ell}(\beta-\tau) G_{i j; \sigma}^{\ell \ell'}(\tau),
\end{aligned}
\end{equation}
where $\ell=A, B$ and $\sigma=\uparrow, \downarrow$.

Based on the Dyson-like expression, the particle-particle interaction vertex $\Lambda_{\sigma \sigma^{'}}^{\ell\ell'}(\mathbf{q})$ relates the susceptibility $\chi_{\sigma \sigma^{'}}^{\ell\ell'} (\mathbf{q})$ to its disconnected part $\bar{\chi}_{\sigma \sigma^{'}}^{\ell\ell'} (\mathbf{q})$:
\begin{equation}\label{eq:magnetic_vertex}
\begin{aligned}
\chi_{\sigma \sigma'}^{\ell\ell'} 
= \bar{\chi}_{\sigma \sigma'}^{\ell\ell'} 
+ \bar{\chi}_{\sigma \sigma_1}^{\ell\ell_1} \Lambda_{\sigma_1 \sigma_2}^{\ell_1\ell_2} \bar{\chi}_{\sigma_2 \sigma'}^{\ell_2 \ell'} + \bar{\chi}_{\sigma \sigma_1}^{\ell\ell_1} \Lambda_{\sigma_1 \sigma_2}^{\ell_1 \ell_2} \bar{\chi}_{\sigma_2 \sigma_3}^{\ell_2\ell_3} \Lambda_{\sigma_3 \sigma_4}^{\ell_3\ell_4} \bar{\chi}_{\sigma_4 \sigma'}^{\ell_4 \ell'} + ...,\\
\end{aligned}
\end{equation}
where we left the $\mathbf{q}$ dependence implicitly and truncate the vertex as it can also be relevant to the relative momentum $\mathbf{k}$ and imaginary time $\tau$. We convert the spin susceptibility tensors to matrices in a new basis $\{A\uparrow,A\downarrow,B\uparrow,B\downarrow \}$ so that the eigenvectors of $\chi(\mathbf{q})$ represent either charge or spin modes: 

when $\mathbf{v}= (a,a,b,b)$, 
\begin{equation}
\begin{aligned}
\mathbf{v} \chi \mathbf{v}^T
= \int d\tau &  \frac{2}{N} \sum_{i,j} e^{i\mathbf{q}\cdot(\mathbf{r_i}-\mathbf{r_j})} \Big\langle (a \hat{n}_{iA}(\tau) + b \hat{n}_{iB}(\tau))
(a \hat{n}_{jA} + b \hat{n}_{jB}) \Big\rangle;
\end{aligned}
\end{equation}

when $\mathbf{v}= (a,-a,b,-b)$,
\begin{equation}
\begin{aligned}
\mathbf{v} \chi \mathbf{v}^T
=  \int d\tau &  \frac{8}{N} \sum_{i,j} e^{i\mathbf{q}\cdot(\mathbf{r_i}-\mathbf{r_j})}  \Big\langle ( a \hat{m}_{iA}(\tau) + b \hat{m}_{iB}(\tau))
(a \hat{m}_{jA} + b \hat{m}_{jB}) \Big\rangle ,\\
\end{aligned}
\end{equation}
where $\hat{n}_{i\ell} = \hat{n}_{i \ell \uparrow} + \hat{n}_{i \ell \downarrow}$ and $\hat{m}_{i\ell} = \frac{1}{2} (\hat{n}_{i \ell \uparrow} - \hat{n}_{i \ell \downarrow})$.

Therefore Eqn.~\ref{eq:magnetic_vertex} can be written in a matrix form as
\begin{equation}
\begin{aligned}
\chi(\mathbf{q}) &= \bar{\chi}(\mathbf{q}) + \bar{\chi}(\mathbf{q}) \Lambda(\mathbf{q}) \bar{\chi}(\mathbf{q}) + \bar{\chi}(\mathbf{q}) \Lambda(\mathbf{q}) \bar{\chi}(\mathbf{q}) \Lambda(\mathbf{q}) \bar{\chi}(\mathbf{q}) + ... \\
&= \frac{\bar{\chi}(\mathbf{q})}{\mathds{1} - \Lambda(\mathbf{q}) \bar{\chi}(\mathbf{q})}.\\
\end{aligned}
\end{equation}

In Fig.~\ref{spin_maxeval_doping}, we plot the maximum eigenvalue of $\Lambda(\mathbf{q})\bar{\chi}(\mathbf{q})$ in the spin mode, labeled as $\Lambda_s(q)\bar{\chi}_s(\mathbf{q})$, under various dopings. When both layers are half-filled, a sharp peak at $(\pi,\pi)$ appears, which is suppressed by $t_{\perp}$ and doping. In the underdoped region, a subleading peak at $(\pi,0)$ reaches its maximum when $t_{\perp} \sim 1.4$. 
We also show the maximum eigenvalue of $\Lambda(\mathbf{q})\bar{\chi}(\mathbf{q})$ in the charge mode, labeled as $\Lambda_c(q)\bar{\chi}_c(\mathbf{q})$, in Fig.~\ref{charge_maxeval_doping}. We observe that the peaks in those plots shift away from $(\pi,\pi)$ as $t_{\perp}$ increases.

\begin{figure*}[thb] 
\centering
\begin{minipage}{1\textwidth}
    \includegraphics[width=0.8\linewidth]{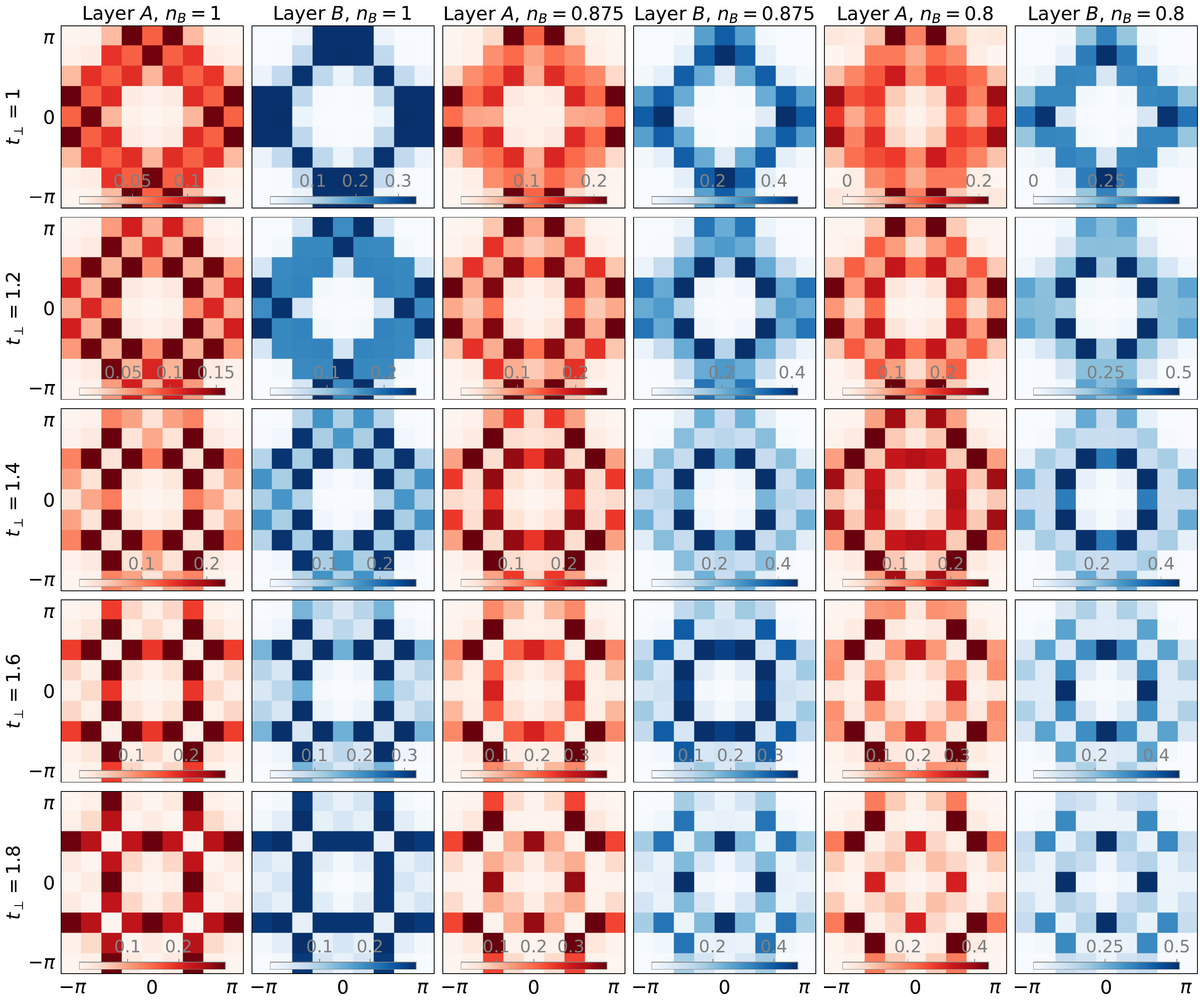}
    \label{FS_dopingB}
\end{minipage}\\
\begin{minipage}{1\textwidth}
    \includegraphics[width=0.8\linewidth]{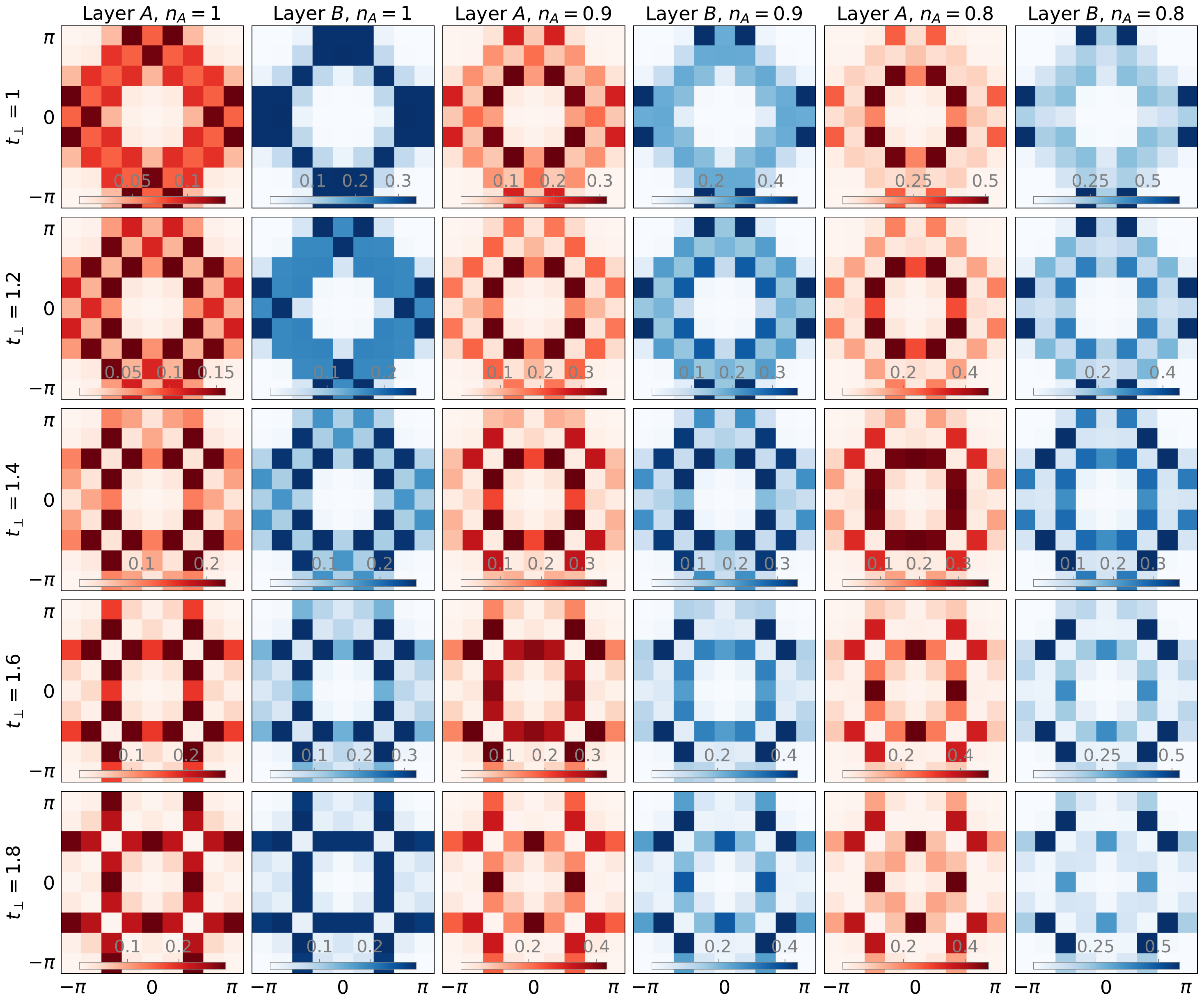}
    \label{FS_dopingA}
\end{minipage}\\
\caption{
\textbf{Fermi surfaces proxy.} The proxy of the zero frequency single-particle spectral weight, $\beta G_{\ell}(\mathbf{k},\tau=\beta/2)/\pi$ with $\ell=A,B$ for various electron densities in (top) layer $B$ and (bottom) layer $A$. The common parameters for all plots are $N=8\times 8$, $\beta=8$, and (top) $n_A=1\pm 0.004$ and (bottom) $n_B=1\pm 0.03$. }
\label{FS_doping}
\end{figure*}

\section{Fermi Surface in Doped Systems\label{appendix:4}}

In Fig.~\ref{FS_doping}, we plot the imaginary time Green function $\beta G_{\ell}(\mathbf{k},\tau=\beta/2)/\pi$ as a proxy for the Fermi surface of each layer ($\ell=A,B$).

\section{DQMC algorithm\label{appendix:5}}

In this study, we perform an unbiased numerical method known as the determinant quantum Monte Carlo (DQMC) method. 
To evaluate the quantum partition function $Z=\Tr(\mathrm{e}^{-\beta \hat{H}})$, we discretize the imaginary time $\hbar\beta$ into $L$ intervals of size $\Delta\tau$, so that $\hbar\beta = L\Delta\tau$. Hence, the partition function can be decomposed using the Suzuki-Trotter decomposition:
\begin{equation}
\mathrm{e}^{-\beta\hat{H}} \approx \prod_{\ell=0}^{L-1} \mathrm{e}^{-\Delta\tau\hat{H}_{t}} \mathrm{e}^{-\Delta\tau\hat{H}_{U}}.
\end{equation}

To transform the quartic interaction term $\hat{H}_{U}$ into a quadratic form, we utilize the Hirsch-Hubbard-Stratonovich transformation by introducing auxiliary field configuration variables $\mathbf{s}$:
\begin{align}
\mathrm{e}^{-\Delta\tau U_{i}  (\hat{n}_{i\uparrow}-\tfrac{1}{2}) (\hat{n}_{i\downarrow}-\tfrac{1}{2})} 
= \mathrm{e}^{-\tfrac{1}{4}\Delta\tau U_{i}} \frac{1}{2}(\mathrm{e}^{\lambda (\hat{n}_{i\uparrow} - \hat{n}_{i\downarrow})} + \mathrm{e}^{-\lambda (\hat{n}_{i\uparrow} - \hat{n}_{i\downarrow})}) 
= \frac{1}{2} \mathrm{e}^{-\tfrac{1}{4}\Delta\tau U_{i}} \sum\limits_{s_{i}=\pm 1} \mathrm{e}^{\lambda s_{i} (\hat{n}_{i\uparrow} - \hat{n}_{i\downarrow})},
\end{align}
where $\cosh(\lambda_{i}) = \mathrm{e}^{\tfrac{1}{2}\Delta\tau U_{i}}$. Here, we successfully rewrite the interaction term in a non-interacting form, which can be explicitly evaluated. With $\mathrm{e}^{-\beta \hat{H}}=\sum_{\mathbf{s}} \rho_{\mathbf{s}}$, the expectation value of a quantum operator can be statistically estimated by sampling a large number of auxiliary field configurations:
\begin{equation}
    \langle\hat{O}\rangle = \frac{\sum_{n} \bra{n} \hat{O} \mathrm{e}^{-\beta \hat{H}} \ket{n}}{\sum_{n} \bra{n} \mathrm{e}^{-\beta \hat{H}} \ket{n}} = \frac{\text{Tr} (\hat{O} \mathrm{e}^{-\beta \hat{H}}) }{\text{Tr} (\mathrm{e}^{-\beta \hat{H}})} = \frac{\sum\limits_{\mathrm{s}}\text{Tr} (\hat{O}\rho_{\mathrm{s}})}{\sum\limits_{\mathrm{s}}\text{Tr} \left(\rho_{\mathrm{s}}\right)} = \frac{\sum\limits_{\mathrm{s}} \langle \hat{O} \rangle_{\mathrm{s}} \text{Tr} \left(\rho_{\mathrm{s}}\right)}{\sum\limits_{\mathrm{s}}\text{Tr} \left(\rho_{\mathrm{s}}\right)}, \;\;\; \text{where}\;\;\langle \hat{O} \rangle_{\mathrm{s}} =\frac{\text{Tr} (\hat{O}\rho_{\mathrm{s}})}{\text{Tr} (\rho_{\mathrm{s}})}.
\end{equation}